\documentclass[
 aps, pra,
 amsmath,amssymb,
 11pt,
 final,
tightenlines,
 twoside,
 twocolumn,
 nofloats,
nofootinbib,
 superscriptaddress,
 ]
{revtex4-2}
\usepackage[T2A]{fontenc}
\usepackage[utf8x]{inputenc}
\usepackage[english, russian]{babel}
\usepackage{graphicx}% Include figure files
\usepackage{epstopdf}
\usepackage{dcolumn}% Align table columns on decimal point
\usepackage{xcolor}
\usepackage{longtable}
\usepackage{afterpage}
\usepackage{amsmath}
\usepackage[version=4,arrows=pgf-filled,mathfontname=mathsf]{mhchem}

\newcommand{\hii}{H\,{II}}

\newcommand{\cii}{С\,{II}}
\newcommand{\CII}{[С\,{II}]}

\newcommand{\OI}{[O\,{\sc{i}}]}  
\newcommand{\oii}{O\,{\sc{ii}}}
\newcommand{\OII}{[O\,{\sc{ii}}]}  
\newcommand{\oiii}{O\,{\sc{iii}}} 
\newcommand{\OIII}{[O\,{\sc{iii}}]} 
\newcommand{\SII}{[S\,{\sc{ii}}]}  
\newcommand{\NII}{[N\,{\sc{ii}}]}  
\newcommand{\trcii}{$^{13}$С\,{II}}
\newcommand{\kms}{km\,s$^{-1}$}

\newcommand{\Halpha}{H$_{\alpha}$}
\newcommand{\Hbeta}{H$_{\beta}$} 
\newcommand{\cmmtr}{cm$^{-3}$}
\newcommand{\nelec}   {$n_{\rm e}$}
\newcommand{\Telec}   {$T_{\rm e}$}
\newcommand{\Brgamma}{Br$_{\gamma}$} 

\newcommand{\AV}{$A_V$} 

\newcommand{\unitsurfbri}{erg~s$^{-1}$~cm$^{-2}$~ster$^{-1}$}

\input{maik.rty}

\setcitestyle{authoryear,round}
\def\squareforqed{\hbox{\rlap{$\sqcap$}$\sqcup$}}

\def\sq{\ifmmode\squareforqed\else{\unskip\nobreak\hfil
\penalty50\hskip1em\null\nobreak\hfil\squareforqed
\parfillskip=0pt\finalhyphendemerits=0\endgraf}\fi}

\def\arcmin{\hbox{$^\prime$}}

\def\arcsec{\hbox{$^{\prime\prime}$}}

\def\utw{\smash{\rlap{\lower5pt\hbox{$\sim$}}}}

\def\udtw{\smash{\rlap{\lower6pt\hbox{$\approx$}}}}

\def\diameter{{\ifmmode\mathchoice
{\ooalign{\hfil\hbox{$\displaystyle/$}\hfil\crcr
{\hbox{$\displaystyle\mathchar"20D$}}}}
{\ooalign{\hfil\hbox{$\textstyle/$}\hfil\crcr
{\hbox{$\textstyle\mathchar"20D$}}}}
{\ooalign{\hfil\hbox{$\scriptstyle/$}\hfil\crcr
{\hbox{$\scriptstyle\mathchar"20D$}}}}
{\ooalign{\hfil\hbox{$\scriptscriptstyle/$}\hfil\crcr
{\hbox{$\scriptscriptstyle\mathchar"20D$}}}}
\else{\ooalign{\hfil/\hfil\crcr\mathhexbox20D}}%
\fi}}

\begin{document}


\title{OPTIMus – a survey of massive star-forming regions at optical, infrared, and millimeter wavelengths}

\author{\firstname{M.~S.}~\surname{Kirsanova} }
\affiliation{Institute of Astronomy, Russian Academy of Sciences, 119017, Pyatnitskaya Street 48, Moscow, Russia}

\author{\firstname{A.~V.}~\surname{Moiseev} }
\affiliation{Special Astrophysical Observatory, Russian Academy of Sciences, 369167, Nizhnij Arkhyz, Karachay-Cherkess Republic, Russia}
\affiliation{Sternberg Astronomical Institute, Lomonosov Moscow State University, 119234, Universitetsky Prospect 13, Moscow, Russia}

\author{\firstname{A.~M.}~\surname{Tatarnikov} }
\affiliation{Sternberg Astronomical Institute, Lomonosov Moscow State University, 119234, Universitetsky Prospect 13, Moscow, Russia}

\author{\firstname{A.~S.}~\surname{Gusev} }
\affiliation{Sternberg Astronomical Institute, Lomonosov Moscow State University, 119234, Universitetsky Prospect 13, Moscow, Russia}

\author{\firstname{A.~D.}~\surname{Yarovova} }
\affiliation{Sternberg Astronomical Institute, Lomonosov Moscow State University, 119234, Universitetsky Prospect 13, Moscow, Russia}

\author{\firstname{D.~Z.}~\surname{Wiebe} }
\affiliation{Institute of Astronomy, Russian Academy of Sciences, 119017, Pyatnitskaya Street 48, Moscow, Russia}

\begin{abstract}

This work presents a description of the scientific goals and objectives of OPTIMus (OPTical, Infrared, Millimeter survey of massive star-forming regions), a survey of massive star-forming regions in the optical, infrared, and millimeter wavelengths. The survey is aimed at constructing a comprehensive characterization of the multicomponent and structurally complex interstellar medium in the vicinity of young massive stars, combining both observational and theoretical aspects. Using multi-wavelength observational data, we will reconstruct the three-dimensional structure and determine the physical parameters of \hii{} regions, photodissociation regions, and the surrounding molecular clouds. The paper describes the observational data obtained with the BTA 6-m and Zeiss-1000 telescopes of the Special Astrophysical Observatory of the Russian Academy of Sciences, the 2.5-m telescope of the Caucasian Mountain Observatory of the Sternberg Astronomical Institute of Moscow State University, and the 20-m telescope of the Onsala Space Observatory.

\end{abstract}

\maketitle

\section{INTRODUCTION}\label{sec:intro}

The dense phase of the interstellar medium (ISM) in galaxies is represented by gas--dust clouds of complex morphology, whose characteristic densities and temperatures span a wide range. Stars and the interstellar medium continuously interact with each other, leading to a variety of effects --from regulating the star formation rate in galaxies to variations in the composition of icy mantles on dust grains in protoplanetary disks. Massive stars make a decisive contribution to the physics of the ISM; despite their small numbers, they influence it through stellar winds, supernova explosions, and intense ultraviolet (UV) and X-ray radiation \citep[see, e.g., the review][]{Zennicker_Yourke_bigreview}. The impact of massive stars on the ISM is most clearly manifested in the optical domain through emission in the \Halpha\ line from regions of ionized hydrogen (\hii), which have been detected in large numbers in the Galactic plane by all-sky surveys, for example,~\cite{SHASSA, IPHAS}.

The theory, whose development began in the mid-twentieth century, shows that shock waves associated with the expansion of \hii{} regions compress neutral interstellar gas and dust, collecting them into dense molecular shells that propagate through the ISM \citep[see][]{Spitzer_1978}. The compression of diffuse gas by intersecting shock waves in the ISM can stimulate the formation of dense filaments, in which stars subsequently form \citep[e.g.][]{1977ApJ...214..725E, Hosokawa_2005}. However, filaments in star-forming molecular clouds can be disrupted by shock waves and UV radiation from massive stars.

In the vicinity of young massive stars that have not yet completely dispersed the gas of their parent molecular clouds, a wide variety of ionization states, physical conditions, and velocities is observed. This is illustrated by the full body of observational data across a broad spectral range -- from radio to ultraviolet wavelengths. Molecular filaments are observed at radio, millimeter and far-infrared (IR) wavelengths \citep[][]{Andre_2010}, ionized hydrogen regions in the optical \citep[][]{Osterbrock06}, and photodissociation regions (PDRs), where physical conditions favor active photodissociation of molecules, in the mid- and near-infrared \citep[][]{Sternberg2014}.

Modern all-sky surveys in the \Halpha{} line clearly demonstrate that ultraviolet radiation from massive stars ubiquitously penetrates inhomogeneities in interstellar gas--dust clouds and ionizes the material not only at their surfaces but also in deeper layers. Spectral lines of atomic ions make it possible to determine key properties of \hii{} regions, such as density and temperature, as well as the type and characteristics of the source of ionization, which may be ultraviolet radiation from a massive star, stellar winds, or shock waves. Despite the significant diversity of spectral types of ionizing stars and the morphologies of the nebulae themselves, only the nearest massive star-forming region in Orion and the \hii{} region surrounding the Trapezium stars have been studied in detail \citep[][]{Tielens_1985, Tielens_1985_ii, Tielens_1993, Abel_2004, 2016ApJ...819..136A, Walmsley_2000, ODell_2009, 2017MNRAS.464.4835O, Rezaei_2020}. The overwhelming majority of other \hii{} regions have been investigated only partially. In addition, the role of expanding \hii{} regions still raises questions in the context of star formation in their surroundings. No examples of spherical \hii{} regions fully embedded in molecular shells with properties corresponding to classical theoretical models have been found yet. However, with the increase of astronomical data over a wide range of wavelengths, it is now possible to conduct detailed studies of the environments of young massive stars and to address the questions outlined above.

In the near future, the launch of the Russian space telescopes \emph{Spektr-UF} \citep[World Space Observatory Ultraviolet, WSO-UV,][]{2016ARep...60....1B} and \emph{Millimetron} \citep[Millimetron Space Observatory,][]{2014PhyU...57.1199K, 2021PhyU...64..386N, 2024PhyU...67..768L} is planned. Given that the \emph{Hubble Space Telescope} is likely to stop operations by that time, \emph{Spektr-UF} will become the importsnt operational spectroscopic facility in the ultraviolet domain among with the Chinese Space Station Survey Telescope \citep[][]{2026SCPMA..6939501C}. \emph{Millimetron} will be the only orbital facility covering the far-infrared and submillimeter wavelength ranges, following the termination of the SOFIA project \citep[][]{2018JAI.....740007R, 2022Natur.605...16W} and the cancellation of the SPICA mission \citep[][]{2018PASA...35...30R, 2020Natur.587..548C}. The preparation of scientific objectives and observational programs for these facilities highlights the importance of studying the structure and physical properties of the multicomponent ISM and star-forming regions. The capability of \emph{Spektr-UF} to observe ionizing radiation and molecular H$_2$ lines in the ultraviolet range will provide unique opportunity to investigate the link between elementary physical processes and the large-scale structure of the ISM. In the context of the planned launch of \emph{Millimetron}, particular emphasis is placed on the modelling and observational study of emission in the \cii{} line at a wavelength of 158~$\mu$m, as one of the spectral bands of the SVR spectrometer is specifically designed for this transition \citep[][]{2025PhyU...68..278K}. Due to the high surface brightness of the \cii{} line emission, it enables efficient mapping of PDRs, as demonstrated, for example, by observations with SOFIA \citep[][]{2020PASP..132j4301S}.

\section{Evolutionary stages of \hii{} regions}\label{sec:stages}

\hii{} regions are among the most easily observable manifestations of the impact of young massive stars on the gas of their parent molecular clouds. Early theoretical models of \hii{} regions assumed spherical symmetry, an ``instantaneous'' onset of ionizing radiation within a neutral gas cloud, and a homogeneous neutral medium, in which the expansion of the ionized region, although decelerated, was limited only by the lifetime of the massive star \citep[][]{Stromgren_1939, 1965ApJ...142.1120M, Spitzer_1978}.

\begin{figure*}
    \centering
    \includegraphics[width=17cm]{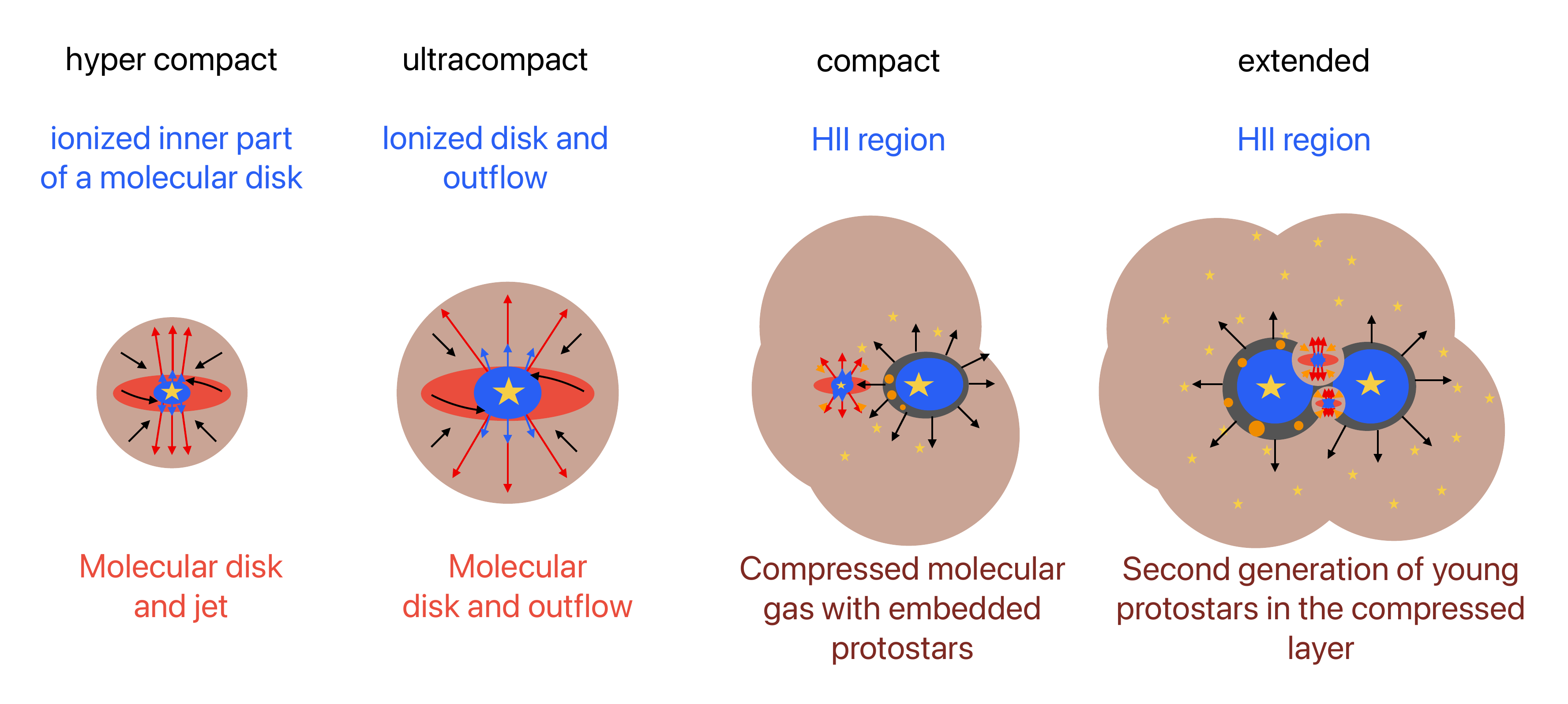}
    \caption{Structure of massive star-forming regions and the morphology of ionized regions. In the schematic, ionized gas is shown in blue, the circumstellar disk in red, and the molecular cloud in brown. The asterisk symbolically represents either a single massive star or a stellar cluster, where multiple stars may contribute to the ionization of the surrounding material. During the hypercompact and ultracompact stages, a molecular disk with outflows (red and brown arrows) is observed around the massive star, along with a collapsing outer envelope. Ionized components of the outflows are indicated by blue arrows. At later stages, a dense molecular shell (dark brown) can be located at distances ranging from 0.1~pc up to several parsecs from the ionizing star. Jeans instability within the dense shell may lead to the formation of a subsequent generation of stars, including massive ones (orange circles oriented along the shell).}
    \label{fig:development_of_hii_regions}
\end{figure*}

The evolutionary stages of \hii{} regions currently identified primarily on the basis of observational studies are schematically illustrated in Fig.~\ref{fig:development_of_hii_regions}. While compact ($d \sim 0.3$~pc) and extended ($d \sim$ a few pc) \hii{} regions are observed at optical wavelengths, hypercompact and ultracompact \hii{} regions ($d \sim 10^{-3}$ and $d \sim 10^{-2}$--$10^{-1}$~pc, respectively) are detected only at millimeter and submillimeter wavelengths (either in the continuum or in radio recombination lines), as they are deeply embedded within molecular clouds \citep[e.g.][]{2005IAUS..227..111K}. Due to the small angular sizes, information on the structure and physical properties of these regions is currently obtained from interferometric observations with facilities such as ALMA and NOEMA; however, these data remain fragmentary and require validation on statistically significant samples. Among the hypercompact \hii{} regions, G24.78+0.08~A1 has been studied in detail. It represents the ionized inner part (up to 500~au from the star) of a molecular disk with a radius of about 4000~au, which hosts a high-velocity jet and is embedded in a collapsing envelope~\citep{Beltran_2007, 2021A&A...650A.142M}. Hypercompact \hii{} regions are characterized by an emission measure ($EM$) of $EM \leq 10^9$~pc~cm$^{-6}$ and an electron density $n_{\rm e} \leq 10^6$~cm$^{-3}$. The structure of the ionized region at the ultracompact stage also appears to consist of an ionized disk and outflow, but they are less bright in the submillimeter range ($EM \leq 10^7$~pc~cm$^{-6}$) and less dense ($n_{\rm e} \leq 10^4$~cm$^{-3}$). Analysis of the radio recombination line widths suggests that the ionization and the growth of the ionized region at these stages are not driven by photoprocesses and/or the thermal expansion of hot gas, as in later stages, but rather by the ionization of gas by a shock wave from a disk wind~\citep[e.g.,][]{Beltran_2007, 2016ApJ...818...52T} and the review \cite{2018A&ARv..26....3A}.

At the early stage of a compact \hii{} region, its mass and size continue to increase until the rates of hydrogen ionization and recombination become equal at a certain distance from the star, resulting in the formation of the so-called Str\"omgren sphere \citep{Stromgren_1939}. The subsequent expansion of the \hii{} region is driven by the thermal pressure difference between the hot ionized gas and the cold neutral medium. The expansion is accompanied by a shock wave propagating ahead of the ionization front into the surrounding gas--dust material, either atomic or molecular. Between the hot ionized gas and the cold neutral cloud compressed by the shock wave, a dense photodissociation region is formed (hereafter PDR; see, e.g., \citealt{Tielens_1985, Tielens_1985_ii, Tielens_1993, Sternberg1995, 1999RvMP...71..173H}). PDRs are irradiated by ultraviolet photons of moderate energy ($h\nu \leq 11$~eV), which are sufficient to ionize heavy elements such as carbon, sulfur, and silicon. This provides a powerful diagnostic of the impact of massive stars on molecular clouds through observations of spectral lines of ions and light hydride molecules \citep{2016ARA&A..54..181G}. PDRs are of particular interest as transition regions in which the transfer of energy and momentum from ionized regions to the molecular gas can be investigated.

Despite the long history of observational studies, direct evidence of the expansion of compressed material in PDRs and molecular shells remains sparse and allows for ambiguous interpretation. Expanding \hii{} regions and associated expanding atomic shells have been identified through spectral-line analysis; see, for example, studies based on observations in the ionized carbon \CII{} line \citep[][]{2019Natur.565..618P, 2020MNRAS.497.2651K}. It has been suggested that some molecular shells may also expand \citep{2011IAUS..270..239D, schneider18, Mookerjea2019, 2019MNRAS.486.2449S, 2025ApJ...990...30F}. However, dense molecular shells exhibiting signatures of shock waves and expanding around \hii{} regions have not yet been observationally confirmed.

The study of the structure of multicomponent ISM provides an important contribution to theoretical models of triggered star formation and its connection to molecular filaments and \hii{} regions. \citet{2012MNRAS.421..408T, 2016ApJ...825..142K} found an enhanced occurrence rate of massive young stellar objects at the peripheries of \hii{} regions. They estimated that the formation of approximately 30\% of all massive stars in the Galaxy may be induced by the accumulation and compression of gas at the edges of \hii{} regions due to the propagation of shock waves.

Although the first models of triggered star formation were proposed as early as the late 1970s \citep[e.g.,][]{1977ApJ...214..725E, 1999AJ....117.2381P}, their observational verification remains challenging and allows for ambiguous interpretations of the data. For instance, using the spatial proximity of ionized regions and clusters of protostars embedded in their parental molecular gas as evidence for star formation triggered by expanding shells and \hii{} regions inevitably leads to false positive associations \citep[][]{2015MNRAS.450.1199D}. On the other hand, the propagation of star formation waves in nearby galaxies on spatial scales ranging from several tens of parsecs up to $0.2-1$~kpc (depending on the galaxy) has been established even after accounting for projection effects \citep{2017ApJ...842...25G, 2019MNRAS.488.3045G}. The same studies demonstrated the dominant role of turbulence in driving the spread of star formation over these scales. However, when one attempts to identify shock waves produced by expanding shells directly and to separate protostars associated with the shock front from the rest of the young population within molecular filaments, the significant impact of projection effects on the results becomes evident \citep[e.g.,][]{2023MNRAS.520..751K}. The \emph{Spitzer} Space Telescope revealed thousands of neutral gas shells surrounding young massive stars \citep[][]{2019MNRAS.488.1141J}, yet the geometry and gas kinematics of these shells are inconsistent with theoretical expectations, even in cases where protostellar infrared sources are present within the shells \citep[e.g.,][see also \citet{Anderson_2015, 2019MNRAS.488.5641K}]{2010ApJ...709..791B}.

The difficulties in studying triggered star formation may be related to the complex structure of shells and their specific velocity fields. In Fig.~\ref{fig:development_of_hii_regions}, the structure of compact and extended \hii{} regions is shown in a simplified manner; however, in reality it is significantly more complex. First, neutral molecular shells surrounding extended \hii{} regions are inhomogeneous and exhibit a clumpy morphology, which complicates—or even makes impossible the detection of their expansion as a coherent structure using position--velocity diagrams \citep[e.g.,][]{2012AJ....144..173D, Anderson_2015, 2015A&A...582A...1D, Trevino-Morales2016}. A typical manifestation of the inhomogeneity in the vicinity of \hii{} regions is the presence of dark globules illuminated along their edges by ultraviolet radiation from nearby hot stars \citep{2011ApJ...741....4F, 2013A&A...559A..31B, 2019AJ....157..112P}. Second, according to theoretical expectations, the expansion velocity of extended \hii{} regions with ages of $0.5$--$1$~Myr embedded in molecular clouds is of the order of $\sim 1$~\kms{} \citep[see][]{Hosokawa_2006, Zavagno_2007, Kirsanova_2009}. This value is lower than the typical velocity dispersions observed in the molecular gas of massive star-forming regions, as demonstrated, for example, by \citet{1987ApJS...63..821S} and in the ATLASGAL survey of massive star-forming regions \citep{2018A&A...609A.125W, 2018A&A...619A.166M}, and than the propagation speed of star formation waves in most nearby galaxies as well \citep{2017ApJ...842...25G, 2019MNRAS.488.3045G}. In such cases, indirect approaches may be more effective. For instance, multiwavelength observations enabled the authors of \citet{2014ApJ...795..121L} to estimate the pressure exerted by ionized gas on the surrounding neutral shells and to conclude that the shell expansion is driven by the pressure difference between the hot and cold media, rather than by stellar winds or radiation pressure.

The \hii{} regions studied in the OPTIMus survey belong to the compact and extended classes shown on the right-hand side of Fig.~\ref{fig:development_of_hii_regions}.

\section{Structure and properties of ISM in star-forming regions}\label{sec:layers}

\subsection*{From ionized regions to molecular clouds}

The most important source of information on the physical state of ionized gas in star-forming regions is optical observations in emission lines. Key parameters such as the electron temperature, electron density, and interstellar extinction are derived from emission lines of ionized atoms. To understand the spatial structure of star-forming regions, it is essential to obtain two-dimensional distributions of these parameters in the plane of the sky. Technically, this is challenging because classical long-slit spectroscopy samples ionization properties only along a single direction. Integral-field spectroscopy provides spectra for each spatial element within the field of view; however, for modern spectrographs this field typically does not exceed one arcminute \citep[e.g.,][]{2015A&A...582A.114W}. This is insufficient for studies of nearby Galactic \hii{} regions. Consequently, investigations of kinematics and ionization parameters over large fields of view rely on Fourier transform spectrographs, such as SITELLE \citep{Drissen2019}, or on spectrometers based on scanning Fabry--Pérot interferometers \citep[FPIs;][]{2021AstBu..76..316M}, as well as on tunable-filter observations employing Fabry-Pérot interferometers \citep{mangal}. These are precisely the techniques planned to be used in the OPTIMus survey. Optical studies of the internal structure of \hii{} regions have so far focused mainly on detailed investigations of the Orion Nebula \citep[e.g.,][]{Abel_2004, ODell_2009}. In many cases, \hii{} regions have been observed not to characterize their morphology, but rather to reconstruct the metallicity distribution in the Galaxy \citep[e.g.,][]{Esteban18}. Such studies are based on long-slit spectroscopy, and we plan to make use of these results in the analysis of the OPTIMus survey data. In addition, recent studies have begun to investigate small samples of Galactic \hii{} regions using the LAMOST multi-object spectrograph; however, these data do not provide sufficient spatial coverage of our targets \citep[][]{2025AJ....169..257Z}.

In PDRs, physical and chemical processes are governed by the absorption of non-ionizing far-ultraviolet photons. The high intensity of ultraviolet radiation leads to the photodissociation of molecules in the parental cloud, so that hydrogen in PDRs is predominantly in atomic form. At the same time, elements such as carbon, sulfur, and silicon are ionized. The densities and temperatures in PDRs are sufficiently high for active chemical reactions, resulting in pronounced chemical complexity. Molecular shielding allows many molecules to survive in PDRs under specific conditions. Molecular emission lines from PDRs allow the physical conditions to be studied from the photodissociation front to the boundary with cold molecular gas. Emission lines from PDRs therefore provide a powerful tool for tracing the evolution of molecular clouds adjacent to massive stars \citep[][]{2023AJ....165...25P}.

The environments around young massive stars—including \hii{} regions, PDRs, molecular filaments, and the surrounding diffuse gas, appear as excellent laboratories for testing theoretical models of interstellar medium evolution and its connection to star formation. Observed emission in lines and the continuum in star-forming regions originates from elementary processes in the medium, such as excitation of molecules, atoms, and ions, ionization, dissociation, and chemical reactions. The efficiency and outcome of these processes depend on the physical conditions, including temperature, density, radiation field, and velocity field. Consequently, addressing the observational and theoretical description of the complex, multi-component ISM around young massive stars is directly linked to the study of these elementary processes and the assessment of their efficiency. A good example is the study the decreased emission of polycyclic aromatic hydrocarbons (PAHs) in the mid-infrared toward \hii{} regions. Although it is generally accepted that this decrease is caused by the photodissociation of PAHs under strong ultraviolet radiation from young massive stars \citep[e.g.,][]{2013ARep...57..573P, 2023ApJ...944L..16E}, detailed calculations based on microphysics and laboratory data indicate that only small PAHs are efficiently destroyed by UV photons \citep[][]{2022MNRAS.509..800M}, while larger PAHs require alternative mechanisms, such as removal by stellar winds \citep{2023IAUS..362..268K}.

\emph{Spitzer} images at 8~$\mu$m and \emph{WISE} images at 12~$\mu$m of PDRs around \hii{} regions often appear as ring-like or arc-shaped structures. These wavelength ranges correspond to the vibrational modes of polycyclic aromatic hydrocarbons (PAHs) \citep{1984ApJ...277..623S, 1984A&A...137L...5L}, excited by ultraviolet radiation. At longer wavelengths, as observed with the \emph{Herschel} telescope, PDRs are traced by the thermal emission of heated dust grains \citep[see e.g.,][]{Deharveng_2005, Churchwell_2006, 2009ApJ...694..546W, Deharveng_2010, anderson_12, Topchieva_2017}. Based on these observations, several thousand PDRs have been identified. However, unlike molecular line emission, PAH or dust emission does not provide information on the kinematics of the medium.

Among the main cooling lines in PDRs are the fine-structure lines of \OI{} at 63~$\mu$m (tracing hot and dense gas) and \CII{} at 158~$\mu$m (tracing warm, less dense gas with temperatures $\leq 200$~K and densities $\sim$ 100--1000~cm$^{-3}$) \citep{Tielens_1985, 1999RvMP...71..173H, 2006A&A...451..917R}. Only high spectral resolution observations of these lines allow the study of gas kinematics and the physical conditions in the emitting region \citep[e.g.,][]{2019Natur.565..618P}.

The third layer surrounding young massive stars, located beyond the \hii{} regions and PDRs, is composed of molecular filaments of dense gas and dust clouds. These filaments evolve continuously, forming new generations of young stars in their cores. This layer is highly inhomogeneous, which affects the gas dynamics in \hii{} regions and PDRs: hot gas can escape through the filaments at their thinnest points, producing high-velocity outflows. Molecular lines from dense filament cores provide sensitive tracers of local physical conditions.

The layered structure of the ISM described above —- from hot ionized regions to cold molecular filaments in star-forming regions -— is embedded within diffuse gas. In the diffuse gas, molecular hydrogen is also photodissociated by absorption of the background ultraviolet radiation. Studying the diffuse gas allows investigation of subtle effects of molecule formation and photodissociation. On the other hand, its presence complicates the study of PDRs, as it absorbs their emission. Investigating structures of the type "\hii{} region—PDR—molecular filaments" therefore requires simultaneously accounting for the influence of diffuse gas on the observed emission.

\subsection*{On the Geometry of Infrared Ring Nebulae}

Since the late 1970s, with the development of observational techniques and the first large-scale \Halpha{} surveys, it has become clear that \hii{} regions are not spherically symmetric, as initially assumed in early theoretical studies. Instead, blister-like, bipolar, and irregular ionized structures were found. These structures were subsequently modeled by many authors \citep[e.g.,][]{1979ApJ...233...85B, Tenorio-Tagle1979, 1990ApJ...349..126F, 1996ApJ...469..171G, 1998MNRAS.298...33R, 2006ApJS..165..283A, 2017MNRAS.466.4573S}.

While infrared ring nebulae observed with \emph{Spitzer}, \emph{WISE}, and \emph{Herschel} are often associated with three-dimensional spherical shells (either \hii{} regions or wind-blown bubbles) in the literature, these objects, which appear as rings in the plane of the sky, may in fact have a different geometry. This is an important issue, as it can provide insights into the detailed distribution of neutral material in star-forming regions, the interaction of massive stars with their parental molecular clouds, the evolution of dust around massive stars, and the possibility and scale of triggered star formation. Nevertheless, the question of the true geometry of infrared ring nebulae remains far from resolved.

An important study on the geometry of infrared ring nebulae was carried out by \citet{Beaumont_2010}. The authors used observations of the CO(3--2) and HCO$^+$(4--3) molecular lines to investigate the gas morphology around 43 infrared ring nebulae identified with \emph{Spitzer}. They demonstrated that the distribution of neutral material around these structures shows no convincing evidence for the presence of front and back molecular walls of \hii{} regions, which would be expected if the nebulae were three-dimensional shells. Furthermore, the kinematic structure of these shells shows no signs of expansion. They therefore concluded that the studied \hii{} regions are enclosed in molecular rings rather than spherical shells, indicating a flattening of the parental molecular clouds. Although the flattened shape of a molecular cloud becomes apparent when an \hii{} region is observed at the edge as a bipolar structure \citep{schneider18}, the shape of infrared ring nebulae requires further detailed analysis. Nevertheless, it is often implicitly assumed that ring-like nebulae have a three-dimensional geometry \citep[e.g.,][]{2010ApJ...713..592E}.

\subsection*{Three-Dimensional Mapping of Star-Forming Regions}

For some relatively nearby star-forming regions, it is possible to construct three-dimensional maps of the matter distribution based on stellar distances and the amount of extinction affecting their light. For example, this approach was used to study the three-dimensional structure of the solar neighborhood in \citet{Leike_2020}, where distances and extinction were determined from Gaia, 2MASS, Pan-STARRS, and ALLWISE surveys. These results were complemented by \citet{Bialy_2021} with CO molecular emission maps, revealing that the molecular clouds in the Taurus and Perseus star-forming regions form an expanding shell with a diameter of approximately 150~pc. In a later study, \citet{Rezaei_2020} applied the same method to a larger volume around the Sun (up to 600~pc) and examined the Orion starforming complex, identifying shells associated with the individual Orion~A and Orion~B clouds. This work was further extended by \citet{2023ApJ...947...66F} using stellar kinematics from Gaia DR2, enabling the reconstruction of not only the structure but also the star formation history in the Orion molecular clouds. They have shown that Orion~A, Orion~B, and $\lambda$~Orionis belong to an expanding, partially empty gas-dust shell surrounding the OBP-B1 stellar cluster, which apparently led to the disruption of the molecular cloud and the formation of the well-known Barnard’s Loop.

The distance- and extinction-based analysis described above can also be applied beyond star-forming regions. For example, \citet{Eiermann_2024} studied the reflection nebulae IC\,59 and IC\,63, illuminated by the star $\gamma$~Cas. They found that IC\,59 lies closer to the observer than the star, whereas IC\,63 is located behind it. These distance estimates allowed the authors to constrain the ultraviolet radiation field in which the nebulae reside, which is important for constructing photodissociation region models.

\subsection*{Recent Studies of PDRs: The Orion Case}

Recent infrared studies of PDRs have largely focused on the Orion region, as well. The PDRs4All project, carried out with the James Webb Space Telescope (\emph{JWST}), has enabled a detailed investigation of the so-called Orion Bar — a molecular shell near the Trapezium stars, seen nearly edge-on \citep[][]{2022PASP..134e4301B}. \citet[][]{2024A&A...685A..74P} showed that the H$_2$ dissociation front in the Orion Bar exhibits a stepped structure, which is related to the geometry of the object. Among the PAH bands toward the Orion Bar, spatial offsets have been observed, associated with molecular restructuring and the incorporation of atoms other than carbon and hydrogen into the molecules \citep[see also][]{2024A&A...685A..75C}. The \emph{JWST} has a small field of view, therefore such observations, while covering a broad and inaccessible from the ground spectral range, remain spatially limited. \citet{Habart_2022} performed near-infrared observations of the Orion Bar (specifically a $40\times40$\arcsec{} field) with the Keck~II telescope using integral-field spectroscopy. These observations revealed that the PDR is composed of numerous substructures in the form of filaments and globules with typical size of $\sim 10^{-3}$~pc ($\sim 10^2$~au).

Studies by \citet{Guzman_2013, Guzman_2014, 2015ApJ...800L..33G, Cuadrado_2015, Cuadrado2017} also are largely focused on objects in Orion: the Orion Bar PDR and the Horsehead Nebula. ALMA interferometric observations of the Orion Bar have shown that it is a non-stationary object, and explaining the relative positions of the H$_2$ dissociation and CO fronts requires the dynamical modelling \citep[][]{Goicoechea2016}, which was then implemented in \citet{2019MNRAS.486.2525K}. Recently, \citet{2023A&A...677A.152H} demonstrated the merging of the atomic hydrogen ionization front and the H$_2$ dissociation front.

Regarding ground-based infrared spectroscopy of PDRs, two studies have focused on the Orion Bar \citep[][]{Marconi_1998, Walmsley_2000}. These works determined the temperature of molecular hydrogen and examined the relative positions of the ionization and dissociation fronts. Progress in this area has been limited by the technical challenges of the observations and the high demand for time on large telescopes. The OPTIMus project aims to address this existing gap.

\section{Aims and Objectives of the OPTIMus Survey}\label{sec:aims}

The main aim of the OPTIMus survey (OPTical, Infrared, Millimeter survey of massive star-forming regions) is to develop a comprehensive observational and theoretical description of the complex, multi-component interstellar medium surrounding young massive stars of different spectral types.

The main scientific objective of this project, within the context of the problem outlined above, is to reconstruct the spatial structure and determine the physical conditions in \hii{} regions, PDRs, and the surrounding molecular clouds using multiwavelength observational data. The survey aims to cover young massive stars of various spectral types that form \hii{} regions and to study objects whose morphologies differ significantly across optical, infrared, millimeter, and radio images. The project plans to analyse data covering the full range from ultraviolet to far-infrared and millimeter wavelengths.

The study of the structure and properties of the interstellar medium in our Galaxy is also highly relevant because these results are often used to interpret observations of other galaxies, due to the limited angular resolution and sensitivity of available instruments. For instance, the relation between the [\CII] 158~$\mu$m luminosity and the star formation rate has been widely employed to study star formation in galaxies, including the Milky Way, under the assumption that this line is optically thin \citep[][]{2013A&A...550A..57O}. However, later analysis of SOFIA infrared observations, whose spectroscopic sensitivity allowed the detection of the \trcii{} 158~$\mu$m line in Galactic PDRs, demonstrated that the line of the main isotope can be optically thick at gas densities above $10^3$~cm$^{-3}$ \citep[][]{2020A&A...636A..16G}. These observations also confirmed that PDRs and star-forming regions are embedded in diffuse gas from the cold neutral phase of the ISM \citep[e.g.,][]{2020MNRAS.497.2651K, 2022A&A...659A..36K}.

\subsection{Spatial structure of \hii{} regions and the role of stellar winds in their formation}\label{sec:optic}

Observed ionized nebulae deviate significantly from the spherical symmetry that is often assumed in theoretical models of these objects. In the OPTIMus survey, we aim to construct parameter maps of ionized regions, including the spatial distributions of electron temperature, electron density, line-of-sight extinction, and dust properties. Such parameter maps will enable the study of the structure of real objects and the identification of both qualitative and quantitative differences from model \hii{} regions, whose properties can be traced dynamically in numerical simulations.

Studying the structure of \hii{} regions requires identifying the ionizing sources and evaluating the relative importance of ultraviolet radiation and stellar winds in shaping the density distribution of the ionized gas. Аll massive stars are expected to drive stellar winds; however, the properties of these winds remain poorly constrained and are typically inferred only indirectly from stellar spectra. Optical observations of \hii{} regions cannot directly reveal cavities filled by a stellar wind. These structures can instead be revealed through X-ray observations, which remain relatively scarce \citep[e.g. archival \emph{Chandra} data were used in][]{2019Natur.565..618P}. As a result, it is still unclear how often stellar winds are the dominant mechanism responsible for clearing nebulae around young massive stars, and in which cases the nebulae are formed primarily through the absorption of ultraviolet photons by the surrounding gas and dust. The study of spatial distribution of electron density around stars of different spectral types can reveal cavities produced by stellar winds 
\citep[][]{2021SciA....7.9511L}.

\subsection{Molecular Hydrogen in PDRs}\label{sec:infrared}

Molecular hydrogen (H$_2$) is the most abundant molecule in the Universe. However, observations of H$_2$ are challenging because it is a symmetric molecule lacking allowed electric dipole transitions. Only quadrupole and magnetic dipole transitions are possible. The observed infrared lines are precisely of this type \citep{2019A&A...630A..58R}. Dense PDRs at the boundaries of \hii{} regions are among the few environments where H$_2$ can be observed in the near-infrared. The corresponding emission lines arise from cascades of ro-vibrational transitions within the ground electronic state following UV photoexcitation of H$_2$, provided that the absorption of UV photons does not lead to molecular dissociation. The presence of this emission indicates active photoexcitation of molecular hydrogen, involving transitions to excited electronic states. Subsequent decay back to the ground electronic state populates excited ro-vibrational levels of the ground term or leads to dissociation of the H$_2$ molecule (transitions to the vibrational continuum). In addition, near-infrared H$_2$ emission can be produced at gas temperatures above $500$~K, when molecular gas is heated by shocks or stellar winds. Within the OPTIMus survey, we aim to determine the locations of H$_2$ dissociation fronts relative to ionization fronts in the objects of our sample, for which the spatial structure of the ionized gas is reconstructed. We also plan to constrain the physical conditions, namely the density and temperature, in the region between the dissociation and ionization fronts, where hydrogen exists predominantly in atomic form.

\subsection{Three-dimensional Structure of Massive Star-forming Regions and the Impact of \hii{} Regions on Star Formation}\label{sec:structure}

In parallel with the study of the spatial structure of \hii{} regions, we plan to reconstruct the three-dimensional structure of their molecular shells, which have been shown to possess a filamentary morphology following the launch of the \emph{Herschel} space telescope. These filamentary shells surround ionized regions and PDRs in a highly non-uniform manner. The structure of star-forming regions observed in the optical often differs from that seen in infrared and millimeter molecular lines and in the emission from cold dust \citep[][]{Anderson_2015, 2019ApJ...882...11A, 2013ARep...57..573P, 2019MNRAS.488.5641K}. Therefore, inferring the structure of an object from observations in a single wavelength range can be misleading. In addition, molecular emission lines arise in relatively cold gas, and their linewidths, even when turbulent broadening is taken into account, typically do not exceed a few km~s$^{-1}$. As a result, molecular lines originating in filaments provide a powerful tool for probing the gas kinematics in star-forming regions \citep[][]{2018ApJ...853..169D, 2018A&A...619A.166M}. We plan to use our data on molecular lines in the far-infrared and millimeter wavelength ranges, together with archival observations from the \emph{Herschel} space telescope, to reconstruct the structure of molecular filaments surrounding \hii{} regions, where the formation of new generations of stars is often observed. At present, the three-dimensional structure of \hii{} regions and the surrounding material is studied mainly using two-dimensional maps of dust emission or molecular line emission, where in the latter case the line-of-sight velocity of ionized, atomic, or molecular gas is adopted as the third coordinate \citep[see, e.g.,][]{Emprechtinger_2009, 2020A&A...639A...2P, 2022A&A...659A..77B}. In the OPTIMus survey, molecular lines will play a supporting role, as the structure will be determined primarily through the approach described in Section~\ref{sec:optic}.

\section{Objects}\label{sec:objects}

\begin{figure*}
    \centering
    \includegraphics[width=0.3\linewidth]{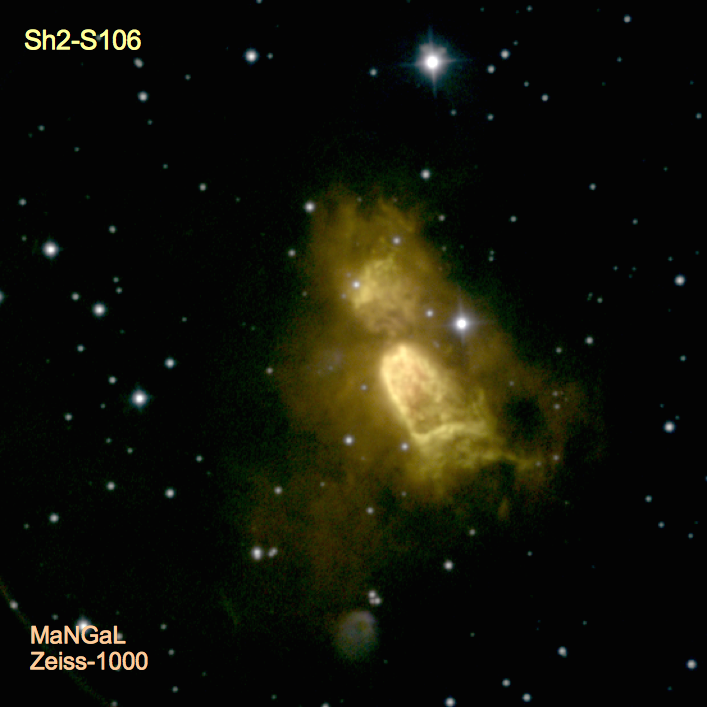}
    \includegraphics[width=0.3\linewidth]{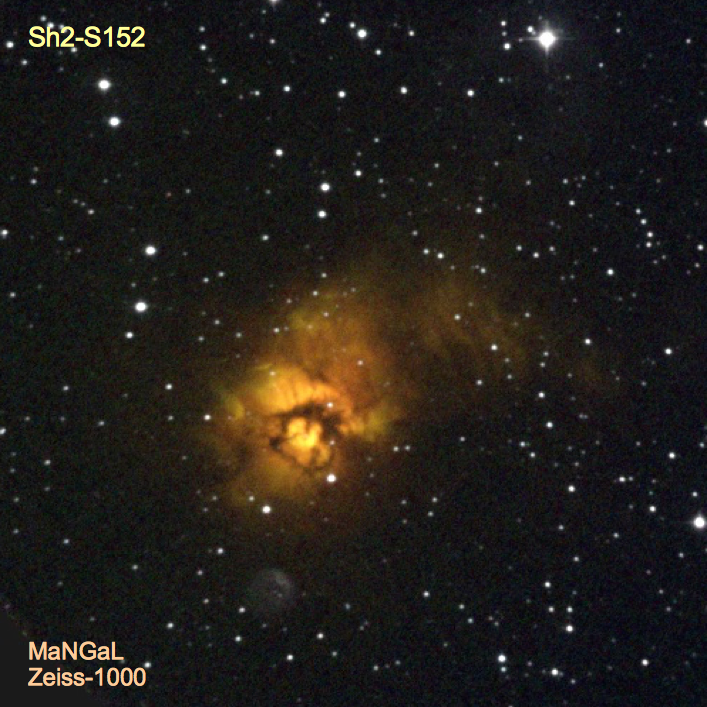}
    \includegraphics[width=0.3\linewidth]{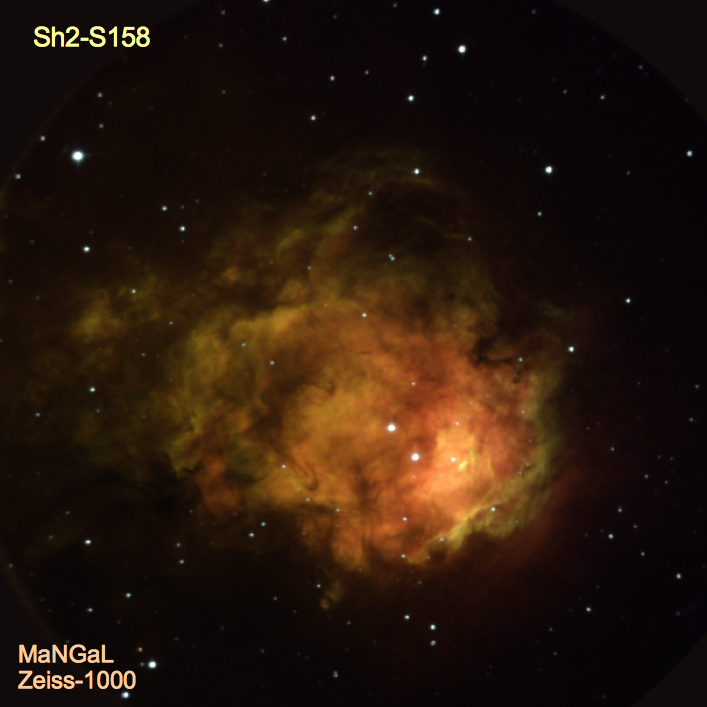}\\
    \includegraphics[width=0.3\linewidth]{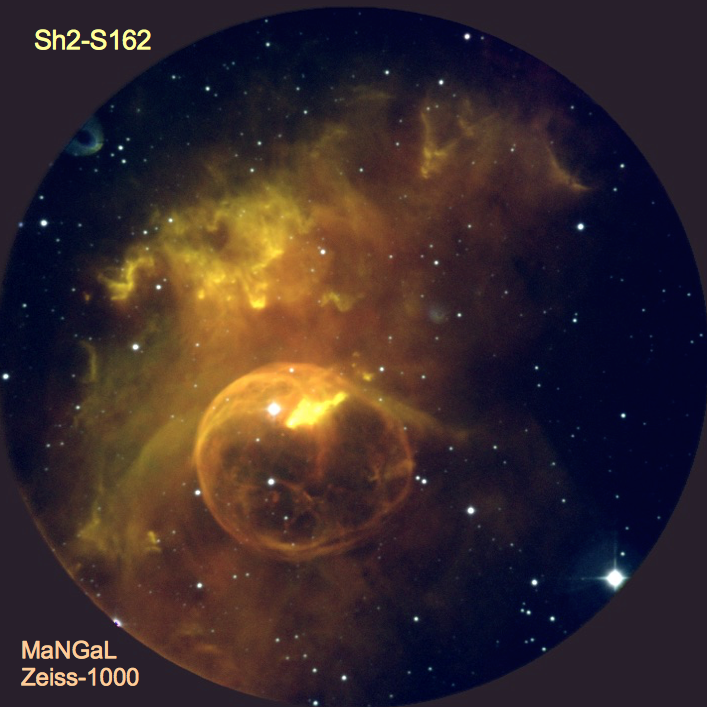}
    \includegraphics[width=0.3\linewidth]{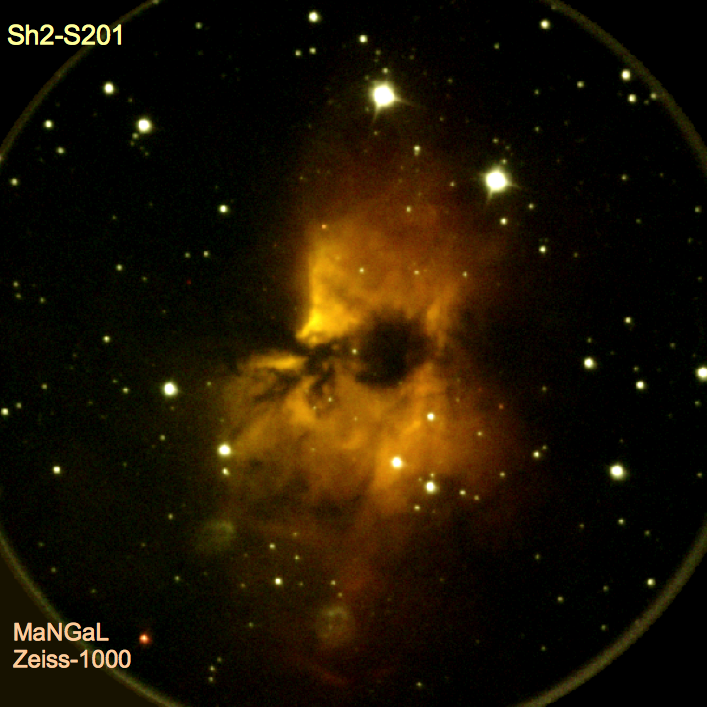}
    \includegraphics[width=0.3\linewidth]{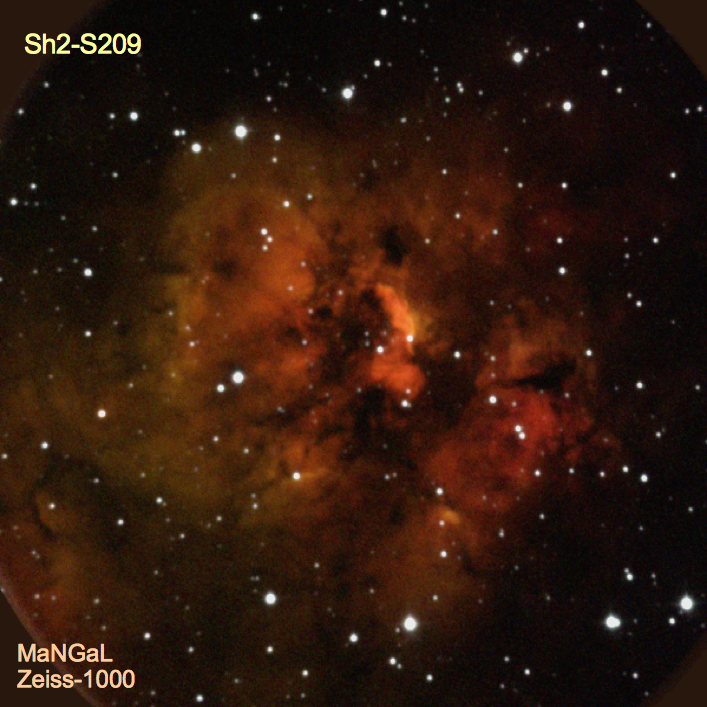}\\
    \includegraphics[width=0.3\linewidth]{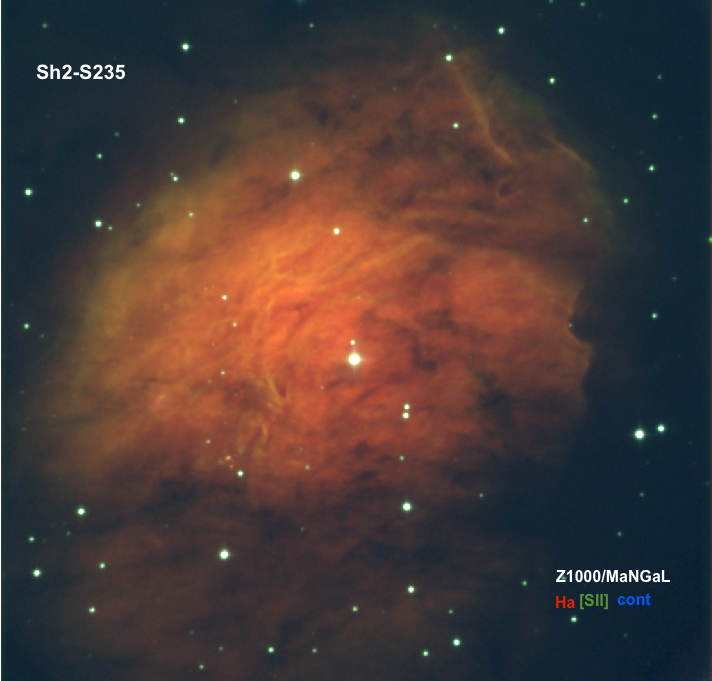}
    \includegraphics[width=0.3\linewidth]{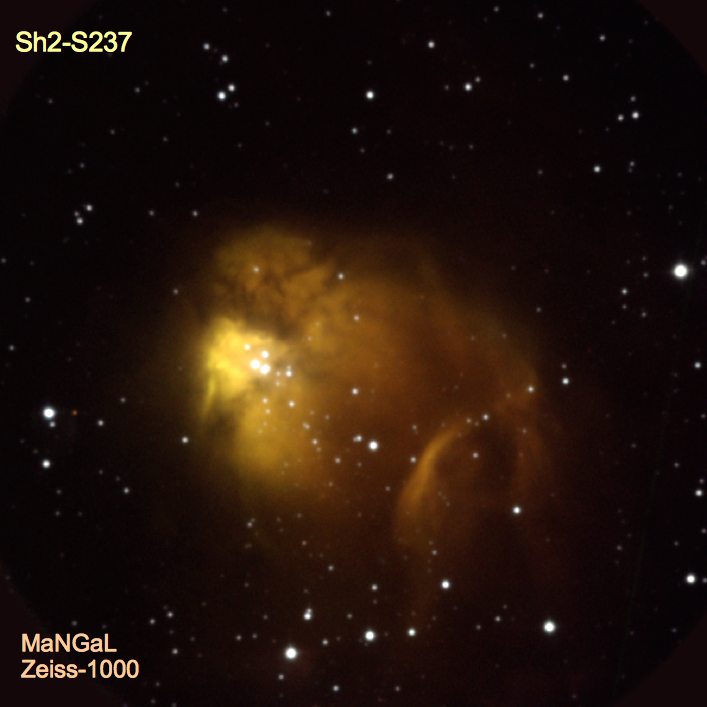}
    \includegraphics[width=0.3\linewidth]{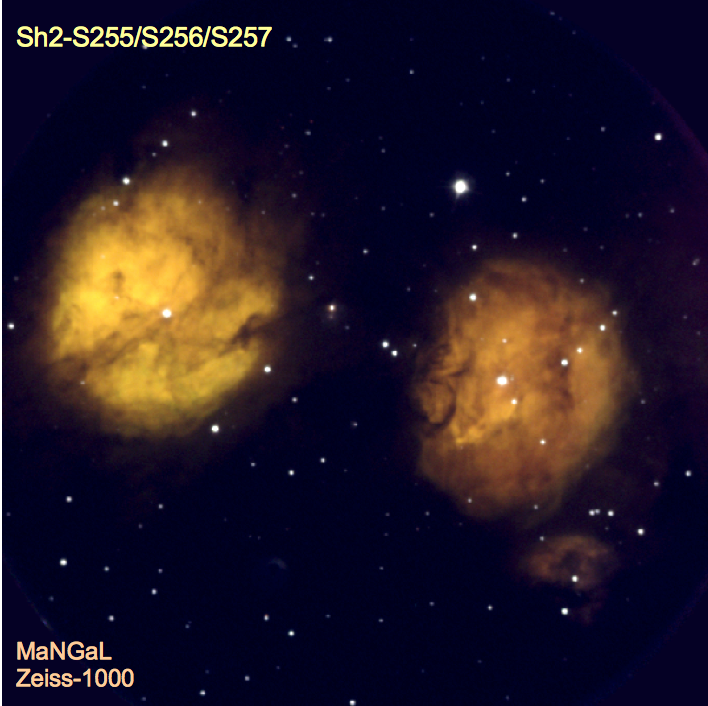}
    \caption{Optical images of \hii{} regions from the OPTIMus survey. The images are taken from the MaNGaL instrument gallery on the website https://www.sao.ru/hq/lsfvo/devices/mangal/}
    \label{fig:objects_optic}
\end{figure*}

OPTIMus survey includes 17 bright \hii{} regions in the northern sky from the catalog of \citet{Sharpless_1959}. Their apparent sizes (up to 10\arcmin) allow imaging and spectroscopy with the the BTA 6-m and Zeiss-1000 telescopes of the Special Astrophysical Observatory of the Russian Academy of Sciences and the 2.5-m telescope of the Caucasian Mountain Observatory of the Sternberg Astronomical Institute of Moscow State University (SAI25 telescope hereafter) (see the description of the methods in Section~\ref{sec:methods}). Some of the selected regions, such as S\,235, S\,237, S\,255, S\,257, and S\,258, are parts of star-forming complexes hosting multiple massive stars. The sample listed in Table~\ref{obs:objects} consists of stars of spectral types from B3~V to O5~V, located at distances of 700–3000~pc from the Sun in the Perseus and Local spiral arms. At present, all optical images and spectra have been obtained (see Fig.~\ref{fig:objects_optic}), and the sample of infrared images and spectra is currently being expanded.

\begin{table*}
\centering
\caption{Parameters of the ionizing stars in the \hii{} regions used in this work. References:
 $^1$\citet{Moffat_1979}, $^2$\citet{Russeil_2007}, $^3$\citet{Chavarria_2008}, $^4$\citet{Gaia_DR3}, $^5$\citet{2008hsf1.book...90H}, $^6$\citet{2020A&A...633A..51Z}, $^7$\cite{2006ApJS..165..338Q} $^8$\cite{2008PASJ...60..961H}, $^9$\cite{2015A&A...580A..83O}, $^{10}$\cite{2011ApJ...738...27B}, $^{11}$\cite{1981A&A....95..206E}, $^{12}$\cite{2011A&A...535A...8R}, $^{13}$\cite{2004ApJ...616.1042O}, $^{14}$\cite{2017A&A...597A..84F}, $^{15}$\cite{1973A&A....25..337G}, $^{16}$\cite{2003ApJ...598.1005F}, $^{17}$\cite{2008PASJ...60..739N}, $^{18}$\cite{2006Sci...311...54X}, $^{19}$\cite{2023ApJ...943..137Y}, $^{20}$\cite{1984A&A...139L...5C}, $^{21}$\cite{2022MNRAS.510.4436M}, $^{22}$\cite{Burns_15}, $^{23}$\cite{2003A&A...397..177B} }
\label{obs:objects} 
\begin{tabular}{|c|c|c|c|c|c|}
\hline
\multirow{2}{*}{Region} & $\alpha$ & $\delta$ & \multirow{2}{*}{Ionizing star} & \multirow{2}{*}{Spectral type} & Distance\\
       & (h:m:s) & (d:m:s) &                 &   & (pc) \\
\hline
S106 & 20:27:26.8 & +37:22:48 & S106IR$^5$             &  late O\,V$^5$ & $1091\pm 54$$^6$       \\
S140 & 22:18:27.8 & +63:13:12 & HD~211880           & B0\,V            & $764\pm27$$^8$        \\
S146 & 22:49:29.0 & +59:54:56 & IRAS~22475+5939  & O7\,V$^{11}$      & $\approx 6200$$^{10}$ \\
S152 & 22:58:41.6 & +58:46:57 & S152~4$^2$             & O8.5\,V$^2$       & $3210\pm210$$^{12}$   \\
S156 & 23:05:10.3 & +60:14:42 & S156~1$^2$             & O8\,V$^2$         & $2870\pm750$$^{2}$    \\
S158 & 23:13:34.4 & +61:30:15 & NGC7538~IRS~6$^{13}$   & O5-O6\,V$^{13}$   & $\approx 2800$$^{13}$ \\
S162 & 23:20:44.5 & +61:11:41 & BD +60$^{\circ}$2522 & O7\,I$^{14}$ & $\approx 4700$$^7$ \\
S165 & 23:39:47.9 & +61:55:42 & BD +61$^{\circ}$2494 & B0~V$^{15}$ & $1900\pm400$$^{16}$ \\
S187 & 01:23:07.3 & +61:51:53 & S187~1$^2$ & B2.5~V$^7$ & $1440\pm260$$^{7}$ \\
S201 & 03:03:17.9 & +60:27:52 & IRAS~02593+6016$^{23}$ & O6-O8\,V$^{17}$ & $\approx 2000$$^{18}$ \\
S209 & 04:11:06.7 & +51:09:44 & ALS 18697$^{21}$ & O9\,III$^{20}$ & $\approx 2500$$^{19}$ \\
S228 & 05:13:25.9  & +37:26:46 & ALS 19710$^{21}$ & O8\,V$^{20}$ & $2560\pm90$$^{21}$ \\
S235 & 05:40:59.4 & +35:50:47 & BD +35$^{\circ}$1201 & O9.5\,V$^{15}$ & $1560^{+90}_{-80}$$^{22}$ \\
S237 & 05:31:26.5  & +34:14:45 & LS V~+34 46 & B2~V$^{21}$ & $2070\pm60$$^{21}$ \\
S255 & 06:13:04.2  & +17:58:41 & LS\,19     & O9.5V$^1$            & $2060^{2181}_{1951}$$^4$  \\
S257 & 06:12:44.2  & +17:59:14 & HD\,253327 & B0.5V$^1$            & $2600\pm160$$^2$  \\
S258 & 06:13:27.6  & +17:55:21 & S258~2$^2$ & B3V$^2$              & $2900\pm510$$^2$ \\
\hline
\end{tabular}
\end{table*}

\section{Observational methods and data analysis}\label{sec:methods}

As the OPTIMus survey combines observations of objects across three broad wavelength ranges -— optical, infrared, and millimeter -— the observational and data analysis methods differ for each of them.

\subsection{Optical spectral range}\label{sec:methods_optics}

\subsubsection*{Extinction}

Determining the physical parameters in ionized regions requires imaging with wide-field filters. In the OPTIMus survey, this is accomplished using a scanning Fabry–Pérot interferometer with a spectral resolution of $\sim13$~\AA. We scan spectral regions around the emission lines \Halpha+\NII, \Hbeta, \OIII{} and \SII. Such observations with medium-sized telescopes became possible in 2017 with the development of the MaNGaL instrument at SAO RAS \citep[][]{mangal}. The planned parameter maps will be constructed based on our observational data obtained with MaNGaL at the Zeiss-1000 telescope of SAO RAS\footnote{https://www.sao.ru/hq/lsfvo/devices/mangal/index.html}. The parameter maps are complemented by spectral “cuts” along selected directions to obtain more precise distributions of the gas parameters as well as dust properties. The spectra have already been obtained with the 6-m BTA telescope of SAO RAS.

By comparing the \Halpha{} and \Hbeta{} line fluxes, we determine the reddening along the line of sight and recover the intrinsic intensity distribution of the objects. To estimate the interstellar extinction at a wavelength $\lambda$: $A_{\lambda} = -2.5 \lg(I_{\lambda}/I_{\lambda,0})$, where $I_{\lambda,0}$ and $I_{\lambda}$ are the intrinsic and observed intensities, we use a reddening law \citep[e.g.,][]{Cardelli89}, in which the mean extinction along the line of sight at a wavelength $\lambda$~($\mu$m) is given by
\begin{equation}
<A(\lambda)/A_V> = a(x) + b(x)/R_V, 
\label{eq:Cardelli_1}
\end{equation}
where $x = \lambda^{-1}$, $A(\lambda)$ is the extinction at wavelength $\lambda$, $A_V$ is the extinction in the V band (central wavelength $\lambda \approx 5500$~\AA), and the coefficients $a(x)$ and $b(x)$ are taken from the study sited above. 

Using the observed line intensities of \Halpha{} and \Hbeta, $I_{\mathrm{H}\alpha}$ and $I_{\mathrm{H}\beta}$, the theoretical values from \cite{Osterbrock06}, and calculating the terms in the denominator of equation~\ref{eq:Cardelli_1}, one can determine $A_V$ as:
\begin{equation}
A_V = \frac{2.5 \lg \left[  (I_{\mathrm{H}\alpha}/I_{\mathrm{H}\beta})/(I_{\mathrm{H}\alpha,0}/I_{\mathrm{H}\beta,0}) \right] }{<A_{\rm H\beta}/A_V>-<A_{\rm H\alpha/}A_V>}.
\end{equation}

In the calculations shown above, the flux ratio $I_{\mathrm{H}\alpha,0}/I_{\mathrm{H}\beta,0}$ is used under the assumption that each transition of an ionized atom to the ground state during recombination immediately produces an ionizing photon, which is then instantly absorbed by a neighboring atom (Case B approximation). This approximation is valid for dense \hii{} regions \citep{Osterbrock06}. Interpolation of the values from Table~4.2 in \cite{Osterbrock06} on a logarithmic scale for a given electron temperature \Telec{} allows one to find the theoretical intensity ratio $I_{\mathrm{H}\alpha,0}/I_{\mathrm{H}\beta,0}$. To estimate $A_V$, one can use the standard value $R_V = 3.1$ for diffuse ISM, or, in HII regions, take into account the deficit of small dust grains and use higher $R_V$ values. Spectra obtained with the 6-m BTA telescope of SAO RAS, which show several lines of the Balmer series, allow us to determine not only the reddening along the slit but also variations in the total-to-selective extinction ratio by analyzing the line ratios $H\gamma/H\beta$, $H\delta/H\beta$, $H\epsilon/H\beta$, and so on. When determining $A_V$, we do not account for the contribution of absorption in the hydrogen Balmer lines. In the spectra, we separate lines formed in the ISM from those formed in stars; in the ISM, absorption in the Balmer lines is negligible.

Using the $A_V$ map to the images of the observed line emission, extinction-corrected images can be obtained using the equation $I(\lambda)=I_\mathrm{obs}(\lambda) \cdot 10^{0.4 A_\lambda}$ in each filter, where the functional form of $A_\lambda/A_V$ is taken from the extinction law \citep{Cardelli89}, and the $A_V$ value for each pixel is derived from the \Halpha/\Hbeta\ line ratio.

\subsubsection*{Electron density}

The electron density of the gas in the ionized regions (\nelec) is determined from the flux ratio of the \SII~$\lambda 6716 / \lambda 6731$ lines in each pixel of the obtained images. For this purpose, we used approximations of the results from the photoionization code CLOUDY \citep{2013RMxAA..49..137F}, as presented in \citet{Proxauf_2014}, specifically equations~\ref{eq:Proxauf14_1} and \ref{eq:Proxauf14_2}. Assuming that the variations in the electron temperature \Telec{} across the nebula are of the order of $\sim 170$~K \citep{Esteban18}, their effect on the \nelec{} values derived from the equations of \citet{Proxauf_2014} is less than 2.5\%. Therefore, when analyzing the spatial variations of \nelec{} across the nebula, possible changes in \Telec{} can be considered negligible. Therefore, the electron density is given by
\begin{equation}
\begin{array}{l}
\log(n_{\rm e}) = 0.0543 {\rm tg}(-3.0553\,R + 2.8506)\,\, + 6.98 - \\10.6905\,R\,\, + 9.9186\,R^2 - 3.5442\,R^3,
\end{array}
 \label{eq:Proxauf14_1}
\end{equation}
where $R$ и $R_\mathrm{obs}^{[{\rm SII}]} (T_{\rm e})$ -- ratio of the line intensities \SII$\lambda 6716/ \lambda 6731$ for $T_{\rm e}=10000$~K, and corrected for the adopted \Telec:
\begin{equation}
\begin{array}{l}
R ={{R_\mathrm{low}^{[{\rm SII}]} (T_{\rm e} = 10 000K)-0.436}\over
{{R_\mathrm{low}^{[{\rm SII}]} (T_{\rm e})-0.436}}}\,\,\left({R_\mathrm{obs}^{[{\rm SII}]} (T_{\rm e})-0.436}\right) \\ + 0.436 \\
\medskip
R_\mathrm{low}^{[{\rm SII}]} (T_{\rm e}) = 1.496 - 0.07442\,\left({T_{\rm e}\over 10^4}\right) + \\ 0.01225\,\left({T_{\rm e}\over 10^4}\right)^2.\\
\end{array}
\label{eq:Proxauf14_2}
\end{equation}

It is well known that numerical models of \hii{} regions around stars with stellar winds produce a shell-like distribution of ionized gas surrounding the star, with a central cavity \citep[][]{1977ApJ...218..377W}, whereas models without winds show a more uniform density distribution. By deriving the distribution of ionized gas from observations, we plan to assess the impact of stellar winds on the dynamics of \hii{} regions.

\subsubsection*{Extent along the line of sight}

To study the geometry of an \hii{} region, its extent along the line of sight ($S$) is estimated for each image pixel using the equation for the emission measure:
\begin{equation}
S = \frac{4\pi I_{\rm H{\alpha}}} {h \nu_{\rm H{\alpha}}} \frac{1}{n_{\rm e}^2 \alpha_{\rm H{\alpha}}^{\rm eff}},
\label{eq:emissionmeasure}
\end{equation}
where $\alpha_{\rm H{\alpha}}^{\rm eff} = 1.17\times{}10^{-13}$~cm$^3$~s$^{-1}$ is the effective hydrogen recombination coefficient for the \Halpha{} line, $I_{\rm H{\alpha}}$ is taken from the original calibrated MaNGaL files, and it is assumed that the proton density $n_{\rm p} \approx n_{\rm e}$, since the ionized gas is composed of hydrogen atoms predominantly.

The calculation of $\alpha_{\rm H{\alpha}}^{\rm eff}$ for a model in which the medium is opaque to transitions to the ground state (i.e., Case B, as opposed to Case A, in which the medium is fully transparent) was carried out using the relation:
\begin{equation}
\alpha_{\rm H{\alpha}}^{\rm eff} = \alpha_{\rm H{\beta}}^{\rm eff} \frac{  \nu_{\rm H{\alpha}}  }{\nu_{\rm H{\beta}}} \frac{ \epsilon_{\rm H{\alpha}} }{ \epsilon_{\rm H{\beta}}  },
\label{eq:effectiverecombcoeff}
\end{equation}
where value of the effective recombination coefficient, $\alpha_{\rm H{\beta}}^{\rm eff} = 3.94\times{}10^{-14}$~cm$^3$~s$^{-1}$, and the emission coefficient ratio, $\epsilon_{\rm H{\alpha}} / \epsilon_{\rm H{\beta}}$, were taken from Table 4.2 in \citet{Osterbrock06} for the adopted electron temperature \Telec.

In the calculations, it is assumed that \nelec{} and \Telec{} remain constant along the line of sight. Furthermore, the contribution of elements other than hydrogen to the free electron density is neglected, i.e., $n_{\rm e} = n_{\rm H^+}$. Regarding the ionization conditions, the abundance of ionized helium, which provides additional free electrons in the \hii{} region, is roughly two orders of magnitude lower than that of H$^+$, while the abundances of other ionized elements are even smaller.

\subsubsection*{Electron temperature}

In the long-slit optical spectra of nebulae ionized by massive stars obtained with the SCORPIO-2 instrument, faint auroral lines are detected in many cases. These lines enable measurements of the electron temperature in the zones of singly and doubly ionized oxygen, \oii{} and \oiii{}.

When the \OIII~$\lambda4363$ line is reliably detected, it becomes possible to measure \Telec{} in the region of doubly ionized oxygen, \oiii{} \citep[the relevant formulae are given in many works; see, e.g.,][]{1992MNRAS.255..325P, Izotov2006, Osterbrock06}. If the [N\,II]~$\lambda5755$ line is observed, \Telec{} can be similarly determined in the region of singly ionized nitrogen, N\,II \citep[e.g.,][]{1969BOTT....5....3P, Izotov2006}. The N\,II region spatially coincides with the region of singly ionized oxygen, \oii{}, and can therefore be characterized by the same value of \Telec.

In some \hii{} regions from the OPTIMus survey, the electron density reaches values of up to $2500~\mathrm{cm^{-3}}$. In this regime, it becomes necessary to account for collisional de-excitation of the energy levels. To compute the electron temperature accurately, we use the Python package \texttt{PyNeb} \citep{2015A&A...573A..42L}, which simultaneously solves for the temperature and density based on multi-level atomic models.

In cases where temperature measurements are available for only one of the ionization zones, the temperatures in the \oii{} or \oiii{} regions can be estimated using the well-known empirical relation $t_{\rm OII} = 0.7\, t_{\rm OIII} + 0.3$ \citep[see][]{Garnett1992}.

Based on the derived electron temperatures, the ionic abundances of O$^{+}$ and O$^{++}$ \citep[see, e.g.,][]{1992MNRAS.255..325P}, as well as those of N$^{+}$ and other ions, can be estimated depending on the set of emission lines detected in the spectra. In the case of oxygen, when no significant contribution from ions with higher ionization stages than O$^{++}$ is expected, the total abundance can be computed as $\mathrm{O/H} = \text{O}^{+}/\text{H}^{+} + \text{O}^{++}/\text{H}^{+}$.

Only in cases of very strong gas heating (specifically, when emission lines with high ionization potentials are detected, such as He\,II~$\lambda4686$) does it become necessary to apply additional corrections for higher ionization stages \citep{Izotov2006}.

\subsubsection*{Ionization diagnostics of \hii{} regions}

Diagnostic maps and diagrams based on spatially resolved spectroscopic data make it possible to characterize the physical conditions in different parts of Galactic nebulae and to identify local processes that remain hidden in integrated measurements. In the proposed survey, we plan both to construct diagnostic diagrams for the interstellar gas sampled by the slit in long-slit spectroscopy and to perform a full-field analysis for the regions for which narrow-band photometry has been obtained with the MaNGaL instrument. The resulting data allow the spatial distributions of various physical parameters across the nebulae to be derived. For example, the \SII~$\lambda6716/\lambda6731$ ratio is sensitive to the electron density and is used to construct maps of the ionized gas density. The \OIII~$\lambda5007/$\OII~$\lambda3727$ ratio traces the distribution of the so-called ionization parameter $U$ (when the \OII{} line cannot be obtained, alternative ratios such as \SII~$\lambda6716,\lambda6731/$\Halpha, \NII~$\lambda6584/$\Halpha, \OIII~$\lambda5007/$\Hbeta{}, and others can be used; see, e.g., \citealt{2010MNRAS.402.1635R}). These indicators are indirectly related to $U$ but are also sensitive to other parameters. In particular, an increase in $U$ is accompanied by an increase in the \OIII/\OII{} ratio and a decrease in the \SII/\Halpha{} and \NII/\Halpha{} ratios, reflecting a higher degree of gas ionization. A high \OIII/\Hbeta{} ratio also highlights regions of enhanced excitation (e.g. near the central stars or shock fronts) and ionization gradients within the nebula. An enhanced \SII~$\lambda6716,\lambda6731/$\Halpha{} ratio may indicate the presence of a shock front; in this case, a spatially coincident high \OI/\Hbeta{} ratio is also expected \citep[e.g.,][]{2012MNRAS.421.3399N,2001A&A...376.1073M}, along with increased velocity dispersion or systematic velocity shifts in the emission lines \citep[e.g.,][]{2018AstBu..73..298O,2021ApJ...915...35L}.

It should be noted that, when analysing spatially resolved data, the demarcation lines of the classical BPT diagrams \citep{BPT, Kewley2001, Kauffmann2003}, which were derived for integrated spectra, as well as the ionized-gas classifications based on them, should be used with caution. Examples of analysis of emission-line ratio diagrams applied to spatially resolved nebulae can be found in \citet{2020MNRAS.493.2238A} and \citet{2024A&A...689A.352K}.

\subsection{Near IR spectral range}\label{sec:methods_IR}

Near-infrared (NIR) observations of OPTIMus survey targets are carried out using the ASTRONIRCAM camera \citep[][]{Nadjip_2017} on SAI25. The camera provides a smaller field of view than MaNGaL, measuring $4.6 \times 4.6$\arcmin. In its grism mode, ASTRONIRCAM delivers low-resolution NIR spectra with a resolving power of $\lambda/\delta \lambda \approx 10^{3.1}$. Imaging is planned in the narrow-band filters \Brgamma{} ($\lambda = 2165$~nm, $\delta \lambda = 21.2$~nm), [Fe\,\textsc{ii}] ($\lambda = 1642$~nm, $\delta \lambda = 26.1$~nm), and H$_2$~1--0~S(1) ($\lambda = 2129$~nm, $\delta \lambda = 46.2$~nm), allowing us to trace the locations of ionisation fronts in atomic hydrogen and dissociation fronts in molecular hydrogen. To facilitate continuum subtraction, additional images are obtained in the narrow-band Kcont filter ($\lambda = 2270$~nm, $\delta \lambda = 39.3$~nm) and the broad-band $H$ filter ($\lambda = 1635$~nm, $\delta \lambda = 291$~nm).

Due to the relatively low extinction affecting these lines along the line of sight from the object to the observer in the near-infrared range \citep[][]{Habart_2022, 2023AstBu..78..372K}, we are able to determine the locations of these fronts with high accuracy. The value of \AV{} derived from the analysis of the optical data allows us to estimate $A_{\mathrm{Br}_{\gamma}}$ and to correct the resulting images and spectra for extinction.

The H$_2$ dissociation fronts can be identified only in images obtained with the H$_2$ filter. By comparing the locations of the fronts and measuring the \Brgamma/H$_2$ line ratio, we plan to constrain the physical conditions—specifically, the gas density and the strength of the ultraviolet radiation field by comparing the observed ratio with theoretical predictions for PDRs \citep[][]{Sternberg2014}.

The near-infrared images will be complemented by long-slit spectra obtained along the directions of the H$_2$ dissociation fronts. Using molecular hydrogen lines detected in the spectra, we will estimate the temperature of the gas excited by the ultraviolet radiation from hot stars, as well as the hydrogen column density along the line of sight. Thus, we will be able to reconstruct the structure of the ionized and atomic components of the interstellar medium surrounding young massive stars.

By comparing the locations of the ionization and dissociation fronts inferred from the observations with those obtained from numerical simulations using the MARION code \citep[][]{Kirsanova_2009, 2019MNRAS.486.2525K}, we can constrain the properties of the medium in which the massive stars formed: whether it is homogeneous or consists of dense clumps embedded in a lower-density gas, or whether the dissociation front itself has a layered structure. Furthermore, by comparing the observed distribution of ionized gas in \hii{} regions with the results of MARION simulations, we will be able to assess the impact of stellar winds on the expansion of the ionized gas.

\subsubsection*{Temperature and column density of H$_2$}

Neglecting the background temperature and assuming local thermodynamic equilibrium, the following expression can be applied, written in a form convenient for constructing a population diagram \citep{Goldsmith1999}:
\begin{equation}
{\rm ln} N = {\rm ln} \frac{N_{\rm u}}{g_{\rm u}} + {\rm ln} Q(T_{\rm ex}) + \frac{E_{\rm u}}{{\rm k} T_{\rm ex}}.
\label{eq:Ntot}
\end{equation}
Here $N$ is the total column density, and $N_{\rm u}/g_{\rm u}$ is the ratio of the column density in the upper level to its statistical weight, where $g_{\rm u} = (2I+1)\times(2J+1)$, $I$ is the nuclear spin, and $J$ is the rotational quantum number of the upper level of the transition. $T_{\rm ex}$ denotes the excitation temperature. The column density in the upper level, $N_{\rm u}$, can be derived from the intensity $I$ of a ro-vibrational H$_2$ line :
\begin{equation}
N_{\rm u} = 4 \pi \frac{I}{A_{\rm ul} \Delta E},
\end{equation}
where $A_{\rm ul}$ is the Einstein coefficient for spontaneous emission and $\Delta E$ is the energy difference between the two levels. By computing the values of $N_{\rm u}/g_{\rm u}$ for several levels, one can construct the so-called population diagram (also known as an excitation diagram), in which ${\rm ln}(N_{\rm u}/g_{\rm u})$ is plotted along the $y$-axis, while $E_{\rm u}/{\rm k}$ is shown on the $x$-axis. In this representation, the slope of the linear fit corresponds to $1/T_{\rm ex}$, whereas the intercept with the $y$-axis yields the total column density $N$. The value of the partition function $Q(T_{\rm ex})$ can be reliably evaluated using the following expression, summing over upper-level energies $E_{\rm u}/{\rm k} \leq 25000$~K \citep[][]{1996AJ....111.2403H}:
\begin{equation}
Z(T_{\rm ex}) = \frac{0.0247 T_{\rm ex} }{1-\exp(-6000/T_{\rm ex})}.
\end{equation}

Since the long-slit spectra obtained with ASTRONIRCAM probe regions of the PDR with varying physical conditions and at different distances from the ionizing stars, we plan to divide the slit into sectors and track how the excitation temperature and the molecular hydrogen column density change with distance.

\subsection{Millimeter wavelength range}\label{sec:methods_mm}

For star formation, the presence of nearby shock sources and strong UV radiation is crucial. High-velocity gas flows can either trigger star formation by compressing diffuse gas or disrupt existing dense star-forming clumps. The same applies to UV radiation. The kinematics of the gas in filaments and their dense clumps are studied through the line-of-sight velocities of molecular lines; the gas temperature and density are simultaneously estimated.

Observations of OPTIMus targets in the millimetre wavelength range were carried out with the 20-m telescope at the Onsala Space Observatory. A dual-polarization 3-mm receiver was used, providing two sidebands with a total bandwidth of $2\times2.5$~GHz \citep[see][]{Belitsky_2015} for a detailed description of the receiver. The spectral resolution was 76~kHz, corresponding to $\sim 0.2$~km~s$^{-1}$, while the beam size varied from $\sim30$ to $40\arcsec$, depending on frequency.

Instead of constructing full maps of the targets in selected lines, the millimetre observations were carried out along the same slit positions as those used in the optical and infrared parts of the survey, in order to ensure consistency between the datasets. The following lines were selected as the primary tracers: the $5_K$--$4_K$ series of CH$_3$CCH and the $2_K-1_K$ series of CH$_3$OH were observed to determine the temperature of the cold molecular gas. The CCH(1--0) lines were used to study the kinematics of photodissociation regions. The pair of lines N$_2$H$^+$(1--0) and N$_2$D$^+$(1--0) was employed to probe the evolutionary stage of star formation in the cold gas. Finally, the CS(2--1) and C$^{34}$S(2--1) lines were used to estimate the column density of the cold molecular gas.

\subsubsection*{Temperature and column density of molecules}

The method for constructing population diagrams, described in the previous section, is used to determine the temperature from the series of lines $5_K-4_K$ of the CH$_3$CCH molecule and $2_K-1_K$ of methanol. In the methanol molecule, both collisional and radiative transitions between K-ladders are allowed, which leads to the establishment of a non-equilibrium level population~\citep[][]{Kalenskii_2016}. However, in the case of CH$_3$CCH, transitions between K-ladders are only collisional. Therefore, we will be able not only to determine the physical parameters of the gas but also to assess the effects of non-equilibrium molecular excitation.

Since only a single line was detected for each molecule, we adopt either the excitation temperature derived from CH$_3$CCH or the dust temperature (see the next section) to estimate the molecular column densities using Eq.~\ref{eq:Ntot}, written in the Rayleigh--Jeans approximation and neglecting the background radiation, where
\begin{equation}\label{eq:Nthin}
\frac{N_{\rm u}} {g_{\rm u}} = \frac{8 \pi {\rm k} \nu^2}{{\rm h} c^3 A_{ul} g_u} \int T dv
\end{equation}
where the statistical weight of the level $g_u = g_J g_K g_I$ is the product of the statistical weights of the upper level --- the quantities of the rotational quantum number $J$, the projection of the total angular momentum $K$, and the spin of the atomic nuclei in the molecule $I$. Given that the spatial extent of the star-forming regions is larger than the telescope beam ($\Theta \approx 30-40$\arcsec), the beam-filling factor is assumed to be 1. The quantity $\int T\,dv$ denotes the integrated line intensity. The value of the partition function $Q(T_{\rm ex})$ and the statistical weight of the level $g_u$ for the molecular transitions are taken from the CDMS database \citep[][]{2001A&A...370L..49M}.

For the pair of lines of the main and rarer isotopologues, CS(2--1) and C$^{34}$S(2--1), the optical depth of the emission, $\tau$, can be estimated as follows: assuming that one of the lines, in this case CS(2--1), is optically thick, while the other, C$^{34}$S(2--1), is optically thin, the optical depth of the optically thick line can be estimated as
\begin{equation}\label{eq:tau}
\frac{T_{\rm thin}}{T_{\rm thick}} = \frac{1-{\rm exp}(-\tau/ r) }{1-{\rm exp}(-\tau) } \approx \frac{\tau / r}{1-{\rm exp}(-\tau) },
\end{equation}
where $T_{\rm thin}$ and $T_{\rm thick}$ can represent either the peak bintensity of the spectral lines or their integrated intensities in the case of complex line profiles, and $r$ is the isotopic abundance ratio. To calculate the optical depth of the CS(2--1) line, we adopt the abundance ratio of the isotopes $^{32}$S to $^{34}$S $r = 22.5$, appropriate for the Galactocentric distances of our objects \citep[$\approx1-2$~kpc][]{Wilson_1999}. Once the optical depth of the transition is known, the excitation temperature can also be derived, either accounting for or neglecting the background radiation (which, in the optically thick case and under LTE conditions, should be close to the gas temperature of the molecular cloud):
\begin{equation}\label{eq:tex}
T_{\rm thick} = ( T_{\rm ex}-T_{\rm bg} ) ( 1-\exp( -\tau) ).
\end{equation}
To correct the column density for optical depth effects, the following expression is used \citep{Frerking1982,Goldsmith1999}:
\begin{equation}\label{eq:goldsmith99}
N_{\rm thick} = N \frac{\tau} {1-{\rm exp}(-\tau)}.
\end{equation}

Therefore, the multi-wavelength observations in the OPTIMus survey enable the determination of physical parameters in the \hii\ region, PDR, and molecular cloud along the same lines of sight in the plane of the sky, allowing the reconstruction of the structure of the star-forming regions.

\subsection{Archival Data Analysis}\label{sec:methods_arch}

\subsubsection{Far-infrared spectral range}

Using archival infrared data, primarily from the Herschel and AKARI telescopes, we estimate the column density of neutral material and compare it with the column density of the material absorbing optical emission. This allows us to determine which of the neutral walls of the \hii{} region (foreground or background) contains more material. Consequently, we can study the three-dimensional structure of the molecular component of the interstellar medium surrounding young massive stars.

Observational data in the far-infrared are obtained from publicly available archives, e.g., ESO\footnote{http://archive.eso.org/cms.html} and IRSA\footnote{https://irsa.ipac.caltech.edu/frontpage/}. For AKARI data, we use infrared emission maps at 90, 140, and 160~$\mu$m (bands WIDE-S, WIDE-L, and N160, respectively), obtained with the Far-Infrared Surveyor (FIS) on board the {\it AKARI} satellite (instrument and satellite descriptions can be found in \cite{Kawada07, Murakami07, Kaneda07}) during the {\it AKARI} Far-Infrared All-Sky Survey (see also \cite{Doi15, Takita15}). From the Herschel telescope archives \citep[see the satellite description in][]{2010A&A...518L...1P}, we use infrared maps at 70, 160, 250, 350, and 500~$\mu$m obtained with the PACS instrument \citep[][]{2010A&A...518L...2P} as part of the Hi-GAL all-sky survey \citep[][]{2010PASP..122..314M}. Before use, all IR data are convolved to a common angular resolution, determined by the longest wavelength.

For each pixel of the IR image, the spectral energy distribution is constructed, and the dust temperature is then determined using a modified blackbody law~\cite{1983QJRAS..24..267H}:
\begin{equation}
I_\lambda = B_\lambda(T)(1-e^{-\tau_\lambda})\approx B_\lambda(T)\tau_\lambda = B_\lambda(T)\Sigma\kappa_\lambda,
\label{eq:dust_emission}
\end{equation}
where $B_\lambda(T)$ is the Planck function, $\Sigma$ is the dust surface density, and $\tau_\lambda$ and $\kappa_\lambda$ are the optical depth and opacity of the dust at a given wavelength, respectively. The dust opacity is defined as
\begin{equation}
\kappa_\lambda=\kappa_0\left(\frac{\lambda_0}{\lambda}\right)^\beta.
\label{eq:dust_emission_kappa}
\end{equation}
where $\beta=-1.59$ and $\kappa_0=0.5$~cm$^2$~g$^{-1}$ at a wavelength of 850~$\mu$m, according to the latest results from the {\it Planck} mission~\cite{Planck_2014}. Fitting the spectral energy distribution in the IR for each pixel allows us to derive maps of the dust temperature.

Surface density of dust grains $\Sigma$ in each pixel is converted into the column density of hydrogen nuclei under the assumption that the gas-to-dust mass ratio is 100:
\begin{equation}
N({\rm HI+H_2}) = \frac{100\Sigma}{1.4m_{\rm H}}.
\label{eq:dust_emission_H_column_density}
\end{equation}

For some objects in the OPTIMus survey, existing maps of dust temperature and hydrogen column density from the ViaLactea survey \citep{Marsh_2017}, based on {\it Herschel} data, can be used. In that work, the spectral energy distribution was fitted using a modified Planck function with $\beta=-1.8$.

For convenience in comparing the column densities of neutral material derived from Balmer line analysis and from dust continuum emission, the hydrogen column density can be related to the extinction via a linear relation, which has been established from ultraviolet observations and reported in the works by \cite{Bohlin_1978, Rachford_2009}:

\begin{equation}\label{eq:NH_AV}
N({\rm HI+H_2}) = 1.87\cdot10^{21} A_V.
\end{equation}

therefore, we have two quantities that can be directly compared: $A_V$ from optical observations and an equivalent extinction, denoted as $A_V^*$, calculated using relation~\ref{eq:NH_AV}.

To estimate the radiation field intensity $G_0$ in each pixel, we use equation 5.44 from \cite{tielensbook}, which relates the dust temperature to the radiation field, varying with distance from the star based on geometric considerations:
\begin{equation}\label{eq:G0}
T_{\rm dust} \simeq 50 \left( \frac{1}{a} \right)^{0.06} \left( \frac{G_0}{10^4} \right)^{1/6},
\end{equation}
where the grain size is given in microns. In the calculations presented, the radius of a typical interstellar dust grain is taken as $a = 0.1~\mu$m. This choice is motivated by the fact that the interstellar grain size distribution, converted by mass fraction, peaks near 0.1~$\mu$m \citep{1994ApJ...422..164K}. In this formula, $G_0$ is expressed in units of the mean interstellar radiation field in the solar neighborhood.

\subsubsection{X-ray and Ultraviolet wavelengths}

To investigate how stellar winds affect the structure of \hii{} regions, it is necessary to use the X-ray observational data, for example from the ROSAT \citep[][]{1982AdSpR...2d.241T}, Chandra \citep[][]{2000SPIE.4012....2W}, or Spektr-RG telescopes \citep{2021A&A...656A.132S}.

To estimate the surface brightness of an extended source in the X-ray band, we use the formula for the bremsstrahlung coefficient \citep[equation 10.1, p.~93 in][]{drainebook}, which is produced in the compressed stellar wind region between the ionizing star and the \hii{} region:
\begin{equation}
j_{\rm ff, \nu} = 5.4\cdot10^{-41} g_{\rm ff} \left( \frac{T}{10^4} \right)^{-1/2}\exp\left( \frac{-{\rm h}\nu}{{\rm k}T} \right) Z^2 n_{\rm p} n_{\rm e},
\end{equation}
where $g_{\rm ff} \approx 1$ is the Gaunt factor and $Z$ is the metallicity. Estimating the surface brightness for the eROSITA energy bands \citep[][]{2021A&A...647A...1P}, and assuming $n_{\rm p} = n_{\rm e} = 1$~\cmmtr, $T = 10^6$~K, a distance to the \hii{} region is 2~kpc, and a size $d = 1$~pc, we obtain $I \approx 6 \cdot 10^{-9}$~\unitsurfbri{} for the $0.5-2$~keV band and $3 \cdot 10^{-8}$~\unitsurfbri{} for the $0.5-10$~keV band, corresponding to fluxes of $5 \cdot 10^{-15}$ and $2 \cdot 10^{-14}$~erg~s$^{-1}$~cm$^{-2}$, respectively. These surface brightness values are at the sensitivity limit of eROSITA \citep[][]{2024A&A...682A..35T}, making cavities filled with stellar wind in \hii{} regions difficult to detect with this telescope. The higher sensitivity of ROSAT and Chandra explains the detection of such cavities in the works cited above.

In the ultraviolet range, GALEX images~\citep[][]{2007ApJS..173..682M} are available for some objects, allowing us to identify the ionizing stars and to study the hottest regions within the nebulae.

\section{First Results}\label{sec:firstres}

As can be seen in Fig.~\ref{fig:objects_optic}, the OPTIMus survey includes objects with a wide variety of morphologies. In this section, we briefly present the results of optical observations of the regions S\,235, S\,255, and S\,257, whose shapes resemble the projection of a sphere onto the plane of the sky. The ionized regions are formed around late O- and early B-type stars (see Table~\ref{obs:objects}).

\begin{figure}
    \centering
    \includegraphics[width=9cm]{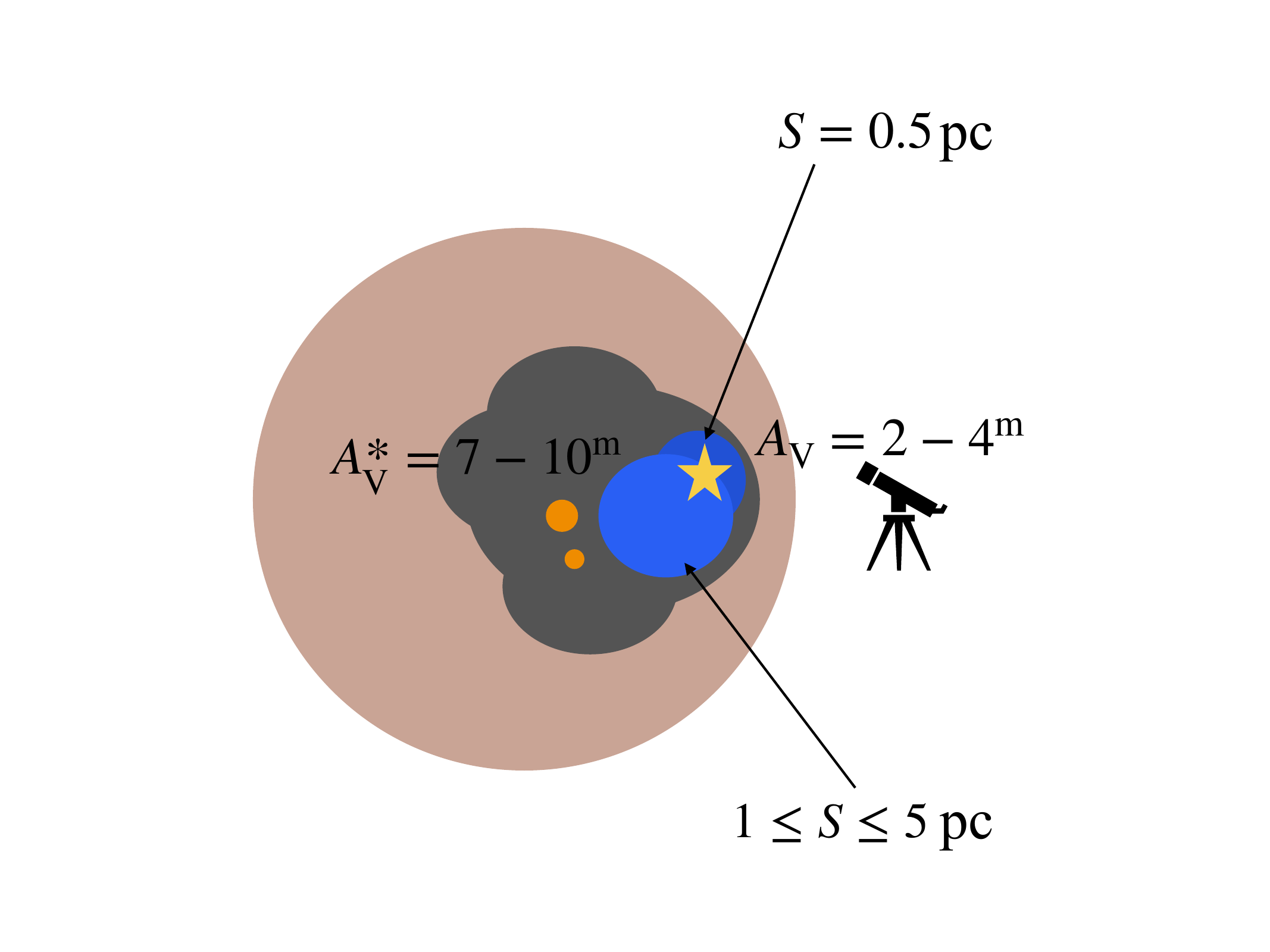}
    \caption{Structure of the S\,235 region. The \hii{} region is shown in blue, with the depth of the color corresponding to higher or lower \nelec. Molecular gas is shown in brown. The ionizing star is marked with a yellow symbol, and young stars embedded in dense molecular gas are marked with orange symbols.}
    \label{fig:example}
\end{figure}

\citet{2020MNRAS.497.1050K} showed that the optical emission of the \hii{} region S\,235 is attenuated by neutral material with $A_V \approx 2$-$4^m$ (see the schematic in Fig.~\ref{fig:example}). The maximum extinction is observed to the southeast of the ionizing star. The direction of maximum extinction coincides with the region of highest electron density, $n_{\rm e} > 300$~\cmmtr. The value of the column density $A_V^* \approx 7$-$10^m$ indicates that the neutral material is predominantly located on the far side rather than the near side. The line-of-sight depth of the \hii{} region ranges from 2~pc in the southwest to more than 10~pc in the northeast.

\begin{figure*}
    \centering
    \includegraphics[width=16cm]{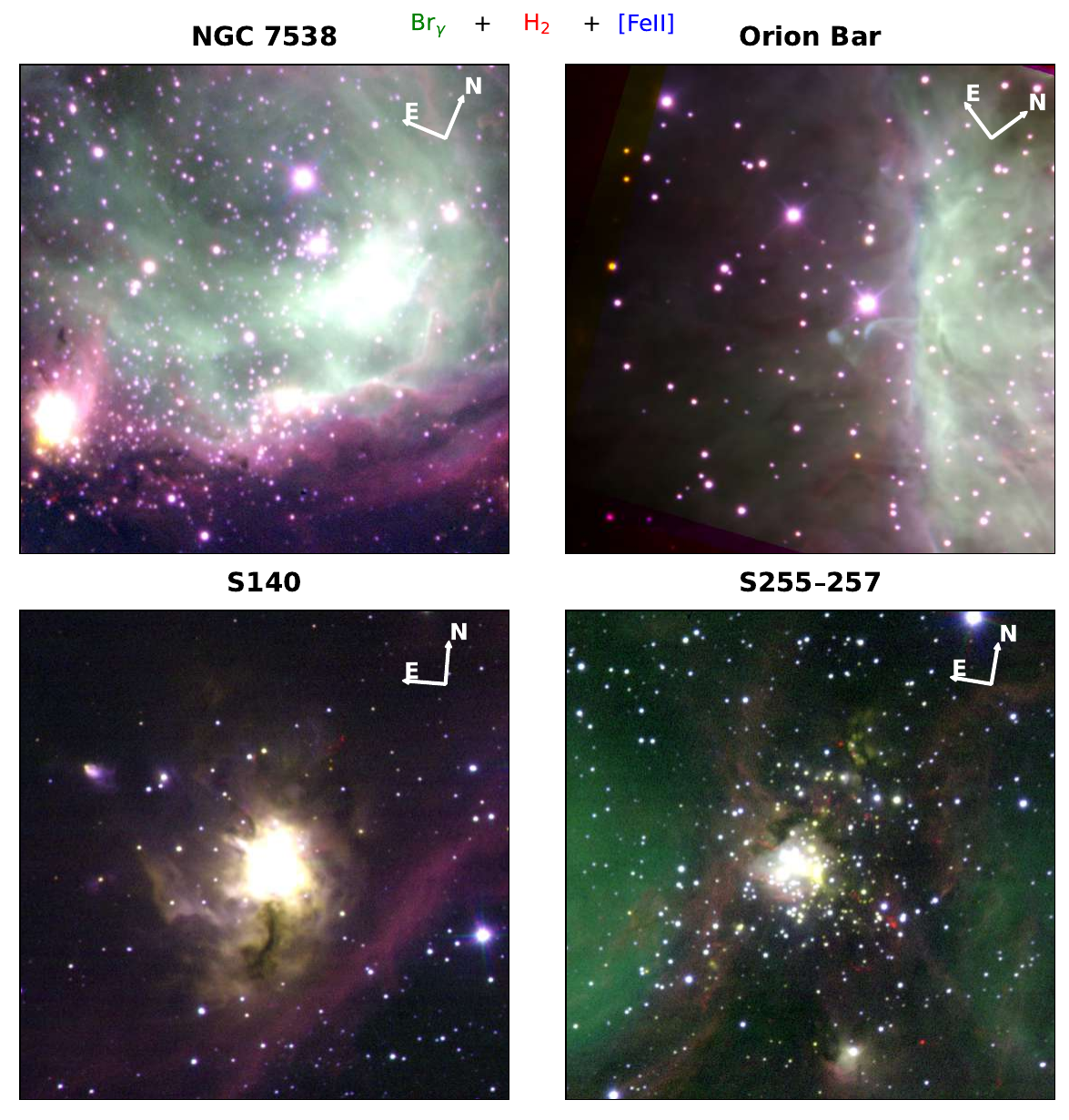}
    \caption{Infrared images of the survey objects in three filters (including continuum) shown in false colors: $\rm Br\gamma$ (green), [Fe\,II] (blue), H$_2$ (red).}
    \label{fig:objects_ir}
\end{figure*}

Later, \citet{2023MNRAS.526.5187K} showed that the optical emission of the \hii{} regions S\,255 and S\,257 is attenuated by absorbing neutral material on the near side, with $2^m \leq A_V \leq 5^m$. In the direction of the dense molecular cloud located between these \hii{} regions, $A_V$ increases due to the presence of the molecular cloud (see Figs.~\ref{obs:objects} and \ref{fig:objects_ir}). The electron density in S\,255 and S\,257 rises from $\sim 100$~\cmmtr{} near the ionizing stars to $\sim 400$~\cmmtr{} at the edge of this cloud. An increase in electron density is also observed along the edges of the ionized regions of S\,255 and S\,257. This increase may be due to the penetration of diffuse UV photons through clumps of dense neutral matter and the subsequent ionization of the latter. Another possible mechanism for forming a partially evacuated shell is the expulsion of ionized gas from the vicinity of young stars by stellar winds.

In three-dimensional space, S\,255 and S\,257 do not resemble each other. In particular, S\,255 is surrounded by dense neutral material on all sides, whereas S\,257 is located at the edge of a molecular cloud and lacks dense material on both the near and far sides. S\,257 is likely a blister-type \hii{} region, as the density of both ionized and neutral gas decreases with distance from the molecular cloud.

Despite the expected high velocities of ionized gas being blown out by stellar winds, \citet{2021ARep...65..488B} estimated an expansion velocity of the molecular part of the shell of S\,255 and S\,257 using the CCH(1--0) as only $\approx 1$~\kms. Increasing the sample of \hii{} regions with measured shell expansion velocities is necessary to study the kinematics of both ionized and neutral gas.

Images of several objects from the OPTIMus survey, previously studied in the near-infrared \citep[][]{2023AstBu..78..372K}, are shown in Fig.~\ref{fig:objects_ir}. To verify the calibration of the infrared photometry of extended objects, we compared our results with the observations of \citet{Habart_2022} and found full agreement in the \Brgamma{} and H$_2$ line fluxes between the SAI25 and Keck~II data.

In the NGC\,7538 region, the ionization and dissociation fronts merge due to the high gas density and the expansion speed of the wind-blown shell. The molecular hydrogen density ranges from $10^{4.5}$ to $10^6$~\cmmtr{} near the H$_2$ dissociation front and drops below $10^4$~\cmmtr{} farther from the star. In contrast, in the S\,255 and S\,257 regions, the projected distance between the ionization and dissociation fronts is about 0.3--0.4~pc, indicating their separation. Comparison with theoretical modeling using the MARION code shows that the ionized gas in S\,255 and S\,257, as well as the dense gas in the PDRs, consists of small dense clumps embedded in a more diffuse continuous medium, which accounts for the observed distance between the fronts. Despite the apparent merging of the fronts in the plane of the sky at our spatial resolution, the HI--H$_2$ transitions in the PDRs of S\,255 and S\,257 are gradual, without sharp boundaries.

\section{Project status}

By the time of publication of this paper, all optical images and spectra have been obtained. In addition, millimeter-wavelength spectra have been acquired with the 20-m telescope at the Onsala Observatory. Infrared images and spectra have been obtained for a subset of the survey objects with the SAI25 telescope, and new observations are currently ongoing. The observational data that have already been analyzed can be found through the links provided in the corresponding articles \citet{2020MNRAS.497.1050K, 2023MNRAS.526.5187K, 2023AstBu..78..372K}.

\section{Conclusion}

As a result of completing all tasks within the OPTIMus project, a comprehensive observational and theoretical characterization of the complex, multi-component interstellar medium around young massive stars will be constructed. The spatial structure and physical conditions in the HII regions, photodissociation regions, and surrounding molecular clouds will be reconstructed based on observational data across a wide range of wavelengths.

The practical significance of the OPTIMus project lies in preparing a research program for \hii{} regions and photodissociation fronts with the upcoming Russian telescopes \emph{Spektr-UF} and \emph{Millimetron}. Within the project, the feasibility of observing specific objects from our sample will be assessed, and a target list for future observations will be compiled.

Thanks to open archives, methods for machine learning and computer analysis of astronomical data are being actively developed. Several international space telescopes are focused on survey projects, providing a basis for detailed studies of individual, either particularly interesting or, conversely, typical objects. The OPTIMus sample can serve as a foundation for future observational programs with \emph{Spektr-UF} and \emph{Millimetron}, which will operate in a targeted mode for individual objects.

\section*{Acknowledgements}
We express our gratitude to N.~N. Chugai, O.~V. Egorov, and the anonymous reviewer of this article for their remarks and comments, which allowed us to make the text more precise and clear.

\section*{Funding}

The formulation of the aims and objectives and the scientific analysis of the obtained data were carried out within the framework of the state assignment of INASAN.

Observations with the SAO RAS telescopes are carried out with the support of the Ministry of Science and Higher Education of the Russian Federation. The upgrade of the instrumentation is implemented within the framework of the national project ``Science and Universities''.

Observations with the 2.5-m telescope of the Caucasus Mountain Observatory of the Sternberg Astronomical Institute, Lomonosov Moscow State University, and the data reduction were carried out within the framework of the state assignment of Lomonosov Moscow State University. This work was performed using equipment purchased with funds from the Moscow University Development Program.

Yarovova A.D. was supported by the BASIS Foundation, grant 25-1-2-88.

\section*{Conflict of Interest}
The authors declare that they have no conflict of interest.

%\section*{Приложение}\label{sec:add}
%\textcolor{red}{оно у нас отсутствует, может убрать заголовок?}

\bibliographystyle{aspb1}
\bibliography{OPTIMus}

\end{document}